\documentclass[runningheads]{svmult}

\usepackage{epsf}   %
\usepackage{psfig}   %
\usepackage{amssymb}   %
\usepackage{makeidx}   
\usepackage{graphicx}  
\usepackage{subeqnar}  
\usepackage{multicol}  
\usepackage{cropmark} 
\usepackage{physprbb}  
\def\fig #1, #2, #3 {
  \smallskip
  \centerline{\psfig{figure=#1,height=#2 in,width=#3 in}} }

\def\\{\hfill\break}

\def\ifm#1{\relax\ifmmode#1\else$\mathsurround=0pt #1$\fi}
\def\kms{\ifmmode\,{\rm km}\,{\rm s}^{-1}\else km$\,$s$^{-1}$\fi} 
\def\hmpc{\,h\ifm{^{-1}}{\rm Mpc}}

\def\ltsima{$\; \buildrel < \over \sim \;$}
\def\lsim{\lower.5ex\hbox{\ltsima}}
\def\gtsima{$\; \buildrel > \over \sim \;$}
\def\gsim{\lower.5ex\hbox{\gtsima}}
 
\def\pmb#1{\setbox0=\hbox{#1}%
 \kern-.025em\copy0\kern-\wd0
 \kern.05em\copy0\kern-\wd0
 \kern-.025em\raise.0433em\box0}
\def\vv{\pmb{$v$}}

\def\v0{\pmb{$0$}}
\def\vnabla{\pmb{$\nabla$}}
\def\div{\vnabla\!\cdot\!}
\def\div{\vnabla\!\cdot\!}
\def\rot{\vnabla\!\times\!}

\def\rotv{\rot\vv}
\def\ifm#1{\relax\ifmmode#1\else$\mathsurround=0pt #1$\fi}
\def\kms{\ifmmode\,{\rm km}\,{\rm s}^{-1}\else km$\,$s$^{-1}$\fi}
\def\hmpc{\,h\ifm{^{-1}}{\rm Mpc}}

\begin{document}
\title*{The Quest for the Cosmological Parameters}
\toctitle{Cosmological Parameters}
%
%
\titlerunning{The Quest for the Cosmological Parameters}
%
\author{M. Plionis\inst{1,}\inst{2}}
\authorrunning{M. Plionis}
%
%
\institute{$^1$ Institute of Astronomy \& Astrophysics, National
Observatory of Athens, 15236, Athens, Greece \\
$^2$ Instituto Nacional de Astrofisica, Optica y Electronica,
Apdo. Postal 51 y 216, Puebla, Pue., C.P.72000, Mexico}

\maketitle              

\begin{abstract}
The following review is based on lectures given in the 1$^{st}$ Samos
Cosmology summer school. It presents an attempt to discuss various
issues of current interest in Observational Cosmology, the selection of
which as well as the emphasis given, reflects my own preference and 
biases. 
After presenting some Cosmological basics, for which I was aided by
excellent text-books, I emphasize on attempts to
determine some of the important cosmological parameters; the Hubble
constant, the curvature and total mass content of the Universe. 
The outcome of these very recent studies is that
the {\em concordance} model, that fits the majority of
observations, is that with $\Omega_\mathrm{m}+ \Omega_{\Lambda}=1$,
$\Omega_{\Lambda}\simeq 0.7$, $H_{\circ}\simeq 70$ km s$^{-1}$
Mpc$^{-1}$, $\Omega_\mathrm{B} \simeq 0.04$ and spectral index of
primordial fluctuations, the inflationary value $n\simeq 1$.
I apologise before hand for the
many important works that I have omitted and for the possible
misunderstanding of those presented.
\end{abstract}

\section{Background- Prerequisites}
The main task of Observational Cosmology is to identify which of the
idealized models, that theoretical Cosmologists construct,
relates to the Universe we live in. One may think that 
since we cannot perform experiments and 
study, in a laboratory sense, the Universe as a whole, this is a futile
task. Nature however has been graceful, and through the detailed and
exhaustive analysis of the detected electromagnetic radiation emitted from the
different photon-generating processes, we can do wonders!

Among the many important tasks of Observational Cosmology is
the determination of the total mass-energy density of the Universe,
the rate of its expansion, its age, the amount of
ordinary and exotic matter that it contains, as well as to quantify
in a objective and bias free manner the large-scale
distribution of matter, as traced
by galaxies, clusters of galaxies and AGN's for example.

However, these tasks are not easy to fulfil. 
Subjective (instrumentation, available funds, technological limitations
etc) as well as objective
(observational biases, limitations due to our position in space-time, etc)
difficulties exist. Furthermore, 
we do not know whether the Universe accessible to our observations 
is representative of the whole Universe. A positive answer to this question is 
essential in order to meaningfully compare observations
with theory. Under the assumption that 
the Universe is homogeneous and isotropic (in a statistical 
sense), well separated regions can be viewed as independent realizations of 
the same formation process. Therefore many of such regions constitute an 
ensemble and thus we can employ statistical techniques in our study.

In the classical Big-Bang cosmological framework the Universe indeed
is considered homogeneous and isotropic on the large-scales.
The most general metric satisfying this
assumption, the so-called {\em Cosmological Principle}, 
is the Robertson-Walker metric (cf. \cite{Weinb}, \cite{CLu}), 
\cite{Peac00}):
\begin{equation}\label{eq:rwm}
ds^{2} = c^{2} dt^{2} - R^{2}(t) \left[\frac{dr^{2}}{1 - k r^{2}} +
r^{2}(d\theta^{2} + \sin^{2}\theta d\phi^{2}) \right] 
\end{equation}
where $R(t)$ is the expansion factor, $k$ is a constant,
related to curvature of space and 
($r, \theta, \phi$) are spherical-polar coordinates. 
The main observational 
evidence that supports the choice of this model is:
\begin{itemize}
\item [$\bullet$] The observed expansion of the Universe.
Edwin Hubble found that the redshifts of galaxies are proportional 
to their apparent
magnitudes and assuming that they are equally luminous 
then their redshifts are proportional to their distances: $v \propto  d$.
\item [$\bullet$] The existence of the cosmic 
microwave background (CMB) radiation, interpreted as
the relic radiation from the hot initial phase of the Universe.
\item [$\bullet$] The observed light element abundances 
that this theory correctly predicts.
\end{itemize}
However, the observed 
matter distribution in the Universe is very inhomogeneous
on small-scales. So, what 
evidence do we have supporting the validity of the Cosmological Principle?

Firstly, the {\em Hubble's law} is directly obtained 
if one assumes a homogeneous expansion of the Universe, ie., if a
length $\chi$ is expanded by a factor $R(t)$, then after some time we have
$d = R(t) \chi$. Differentiating we obtain the Hubble law:
\begin{equation}\label{eq:hub}
v = \frac{\dot{R}}{R} \; d = H(t) \; d \;,
\end{equation}
where
$H(t)$, at the present time ($t=0$), is the Hubble constant. Secondly,
observations of distant extragalactic radio sources have shown that 
they are distributed across the sky in a uniform way.  
Other supporting evidence is provided from the decreasing - with
scale - correlations of extragalactic objects and the directional
independence of the correlation function. However,
the most remarkable 
confirmation that the Universe is 
homogeneous and isotropic and also that it evolved from a hot dense past, 
was the discovery of the cosmic microwave background radiation 
and its high degree of isotropy across the sky. 
This radiation has been interpreted as the relic radiation from the time that 
matter decoupled from radiation, which has been freely travelling ever 
since. The high degree of isotropy of the CMB is direct evidence that the 
Universe was highly isotropic at the epoch of decoupling ($z \sim
1100$) to one part in $10^{5}$ on scales from arc-minutes to
$90^{\circ}$, once we subtract a local dipole
anisotropy, attributed to our 
peculiar motion with respect to the rest-frame defined by the CMB.

\subsection{Basic Elements of Dynamical Cosmology}
Out of the many parameters which are essential for accurately
determining the global dynamics and the formation history of the
Universe; three clearly stand out: the Hubble constant,$H_{\circ}$, 
which tells us the expansion rate of the Universe, the matter density 
parameter, $\Omega_\mathrm{m}$, which tells us how much 
matter, being baryonic or exotic, the Universe contains 
and the Cosmological constant, $\Lambda$,
which tells us whether the universe is filled with an extra repulsive
force. The way in which these parameters affect
the cosmological evolution, are determined by the gravitational field 
equations and the assumed equation of state. Below, I will sketch
the basic framework of standard Cosmology, in order to derive the
interrelations of these parameters and how they relate to the global dynamics
of the Universe.

\subsubsection{Friedmann Equations:}
Within the context of a homogeneous and isotropic Universe we can derive
the cosmological evolution equations using either
Einstein's field equations or Newtonian gravity. 
The latter is possible exactly because of the
Cosmological Principle, ie., we can consider any volume element, as small as
necessary for Newtonian gravity to apply, as being representative of the
Universe as a whole. We will derive the evolution equations for the
Newtonian case, using however as an active mass density:
\begin{equation}\label{eq:activ}
\rho^{'} = \rho + 3p/c^2
\end{equation}   
which is given within the framework of general relativity
for a homogeneous and isotropic fluid with density $\rho$ and pressure $p$.

In this case we can derive the evolution equations using the mass 
continuity equation and Newton's equation of the motion of a small
sphere. We have from the homogeneity assumption that $\nabla\rho = 0$,
from isotropy that $\nabla\cdot \vec{v} = 3 H(t) = 3\dot{R}/R$ and
then from $\partial \rho/\partial t + \nabla\cdot(\rho\vec{v}) =0$ we obtain:
\begin{equation}\label{eq:contin}
\dot{\rho} + 3 \left[\rho+\frac{p}{c^2}\right] \frac{\dot{R}}{R} = 0
\end{equation}
Rearranging (\ref{eq:contin}) and using
Newton's equation of motion, $\ddot{R} = -G M/R^{2}$, we have:
\begin{equation}
\frac{d}{dt}\left[\frac{1}{2} \dot{R}^2 - \frac{4 \pi G}{3} \rho R^2 \right] 
= 0
\end{equation}
and integrating we obtain a Newtonian analogue 
of cosmological evolution equation:
\begin{equation}\label{eq:foundam}
\frac{\dot{R}^2}{R^{2}} -\frac{8 \pi G}{3} \rho  = \frac{\cal C}{R^{2}} 
\end{equation}
where ${\cal C}$ is the constant of integration, which is closely
related to the 
Newtonian energy. We can see this if we rearrange (\ref{eq:foundam}) as:
$\dot{R}^2/2 - G M/R = -{\cal C}/2$.
Within the relativistic formulation, the constant is in effect related
to the curvature of space; ${\cal C}= kc^2$ and the basic equation 
of cosmological evolution, the {\em Friedmann equation}, is written:
\begin{equation}\label{eq:ein1}
\frac{\dot{R}^{2}}{R^{2}} + \frac{k c^{2}}{R^{2}} = \frac{8 \pi G \rho}{3} + 
\frac{\Lambda c^{2}}{3}
\end{equation}
where the cosmological constant ($\Lambda$) term was 
introduced by Einstein (rather {\em ad hoc}) in order to obtain his preferred 
static solution. Given an equation of state, $p = p(\rho)$, we can solve 
for $\rho$ and then (\ref{eq:ein1}) can be integrated to give $R(t)$.  

If we recall Newton's equation of motion, $\ddot{R} = -G M/R^{2}$, and
we use the active density (\ref{eq:activ}) and (\ref{eq:ein1}), we
derive the second important dynamical equation of cosmological
evolution. The correct relativistic form of this equation is:
\begin{equation}\label{eq:ein2}
\frac{2 \ddot{R}}{R} + \frac{\dot{R}^{2}}{R^{2}} + \frac{k c^{2}}{R^{2}} 
= -\frac{8 \pi G p}{c^2} + \Lambda c^{2}
\end{equation}

\subsubsection{Equation of State:}
The question arises of which is the appropriate equation of state for
the expanding Universe. The Universe expands adiabatically\footnote{An 
adiabatic process is defined as that in which there is no heat flow 
and thus the entropy is conserved ($\D S=0$).}, since the symmetry
imposed by the Cosmological principle implies that there is
no net heat flow through any surface. Therefore as it expands, it
cools (\ref{eq:cools}) and since it started
with a very hot phase, it will
be dominated, at different
epochs, by different species of particles having distinct equations of
state. Relativistic particles will dominate the hot phase while
non-relativistic the later cooler phases.

We can specify a unique equation of state $p=p(\rho)$ for all epochs by
parameterizing it according to:
\begin{equation}\label{eq:state}
p = w \langle v^2 \rangle \rho
\end{equation}
where $\langle v^2 \rangle$ is the velocity dispersion of the fluid elements.
If the dominant contribution to the density comes from relativistic 
particles, which have $p=1/3 \rho c^2$ (when $kT \gg m_{\circ}v^{2}$, 
with $v^2 \simeq c^2$), then $w=1/3$. 
If the dominant contribution comes from non-relativistic 
matter (when $kT \sim m_{\circ} v^{2}$ with $v^2 \ll c^2$) then there is
negligible pressure and the dust approximation is excellent
($w=0$).

Therefore inserting (\ref{eq:state}) into the mass continuity equation, 
(\ref{eq:contin}), we obtain:
\begin{equation}\label{eq:rho1}
 \rho \propto R^{-3(1+w)} 
\end{equation}
Armed with (\ref{eq:rho1}) one can now solve the Friedmann equation 
to get the time evolution of $R(t)$, determine the age of the
Universe, etc, in the different Cosmological models.

\subsection{Thermal Beginning of the Universe}
The early universe, where very high densities and temperatures
dominate, can be treated by using fluid thermodynamics.
At very high temperatures, radiation and matter are in thermal
equilibrium, coupled via Thomson scattering with the photons 
dominating over the nucleons ($n_{\gamma}/n_\mathrm{p}\simeq 10^9$). 
Therefore the primordial fluid can be treated as 
radiation-dominated with $p=1/3 \rho c^2 = 1/3 \sigma T^4$ and
from (\ref{eq:rho1}), we obtain:
\begin{equation}\label{eq:cools}
T \propto R^{-1}
\end{equation}
Therefore the temperature of the Universe drops linearly with the
expansion scale factor. Furthermore, it is evident from
(\ref{eq:rho1}), that the 
radiation density drops faster than the mass density and since we know from
measurements that the universe is matter dominated today, then at some epoch 
in the past, say at a redshift $z_\mathrm{eq}$, 
we had $\rho_\mathrm{m} = \rho_\mathrm{rad}$. It is easy to show that
$\rho_\mathrm{r} = \rho_\mathrm{m} R_\mathrm{\circ}/R_\mathrm{eq} 
= (1 + z_\mathrm{eq})\; \rho_\mathrm{m}$ (the
subscript $\circ$ denotes the present epoch)
and using the measured values of $\rho_\mathrm{i}$ we have that:
$$1+z_\mathrm{eq} \simeq 2.3 \times 10^{4} \; h^{2} \; \Omega_\mathrm{m}$$
Therefore the thermal history of the Universe can be divided in two main
eras: a {\em radiation dominated era} ($z\gg z_\mathrm{eq}$) and a
{\em matter dominated era} ($z\ll z_\mathrm{eq}$).
In the radiation dominated era, in which we can neglect 
the curvature and $\Lambda$ terms in Friedmann's equation
(see next section), we have:
 $$R\propto t^{1/2} \;.$$
By differentiating this relation with respect to time and using
(\ref{eq:ein1}) we have:
\begin{equation}
t=\left(\frac{3}{32 \pi G \rho_\gamma} \right)^{1/2} \;.
\end{equation} 
Using $\rho_\mathrm{\gamma} = \pi^2 k_\mathrm{b} T^4/15 h^3 c^5$ 
we finally obtain the important 
relation between cosmic time and the temperature of the Universe in
the radiation dominated era:
\begin{equation}\label{eq:T-t}
T_\mathrm{Kelvin} \simeq 1.3 \times 10^{10}
t_\mathrm{sec}^{-1/2}
\end{equation}
from which it is evident that the Universe at early times was hot enough for
nucleosynthesis to occur, as it had been supposed originally by Gamow. 
The era of nucleosynthesis takes place around $\sim 10^{9}$ K.

\subsubsection{The Cosmic Microwave Background:}
Although the dynamics during the {\em radiation dominated era}
are unaffected by ordinary matter, the
electrons act as a scattering medium of the radiation and thus the
Universe at this epoch is {\em opaque}.
As the Universe cools, $\propto R^{-1}$, electrons bind
electrostatically with protons to form Neutral Hydrogen. Using {\em
Saha's ionization} equation one finds that the
temperature at which the number of free electrons drops significantly
is $T\simeq 3000$ K. 

Therefore when the universe cools at this temperature, the
scattering medium disappears and the radiation freely escapes without
being absorbed or scattered which means that the Universe becomes
transparent. This epoch is called the {\em recombination} epoch.

The existence of the relics of this radiation was predicted by Gamow
and his collaborators in the 1940s. It was subsequently discovered
by Penzias \& Wilson in 1965, while the whole spectrum of this 
radiation was traced to unprecedented accuracy 
by the {\sc COBE} satellite observations.
The CMB possesses a perfect {\em black-body} spectrum with a
mean temperature of $T_\mathrm{\circ} = 2.728 \pm 0.004$ K and it is extremely
isotropic except for a dipole, which is however a local kinematical
effect (due to our motion with respect to the cosmic rest frame defined
by the CMB).
From what redshift does the CMB radiation originate?
From (\ref{eq:cools}) we have that:
$$ \frac{T}{T_\mathrm{\circ}} = \frac{R_\mathrm{\circ}}{R} = 1+z$$
with $T\simeq 3000$ K and $T_\mathrm{\circ} \simeq 2.73$ we get that
$$1+z_\mathrm{rec} \simeq 1100$$
From (\ref{eq:ein1}) and (\ref{eq:rho1}) we have that in the matter
dominated era $R(t) \propto t^\frac{2}{3}$ and thus $z_\mathrm{rec}$
corresponds to a time:
$$t_\mathrm{rec}\simeq 2.8 \times 10^{-5} \; t_{\circ}$$
where $t_{\circ}$ is the present age of the Universe.
Therefore by studying the microwave background 
sky we have direct information from
the Universe when it was as young as $t_\mathrm{rec}$.

\subsubsection{The CMB dipole anisotropy:}
Due to our motion with respect to the isotropic CMB radiation we observe
a dipole in the distribution of the radiation temperature. Although
this has the appearance of a {\em Doppler} effect, in reality four
different effects add up to produce this dipole seen by an observer
moving with a velocity $u$. These four effects are:
\begin{itemize}
\item a Doppler effect that increases the frequency of photons, and
thus the observed energy, seen in the
direction of motion by a Doppler factor $D\equiv  1 + (u/c) \cos \theta$
\item the interval of frequencies increases by the same factor in the
direction of motion, and therefore since $T\propto E/\delta \nu$, the
above two effects cancel out.
\item the moving observer selects in the direction of motion 
relatively more photons by a factor $D$
\item the solid angle in the direction of motion is smaller 
by a factor $D^{-2}$ due to abberation.
\end{itemize}
The net effect is that the moving observer sees an intensity of CMB
radiation $I_\mathrm{mov}= (1 + u/c \cos \theta)^3 I_\mathrm{rest}$.
Due to the adiabatic expansion of the Universe, ($T\propto R^{-1}$), 
the shape of the Planck spectrum:
$$I_\mathrm{\nu} = \frac{4 \pi h \nu^3}{c} 
\left[\exp{\left(\frac{h\nu}{kT}\right)}-1
\right]^{-1}$$
 should be preserved, which then necessarily implies that
$T(\theta)= (1 + u/c \; \cos \theta) T_{0}$ and thus:
\begin{equation}
\frac{\Delta T}{T}=\frac{u}{c} \cos \theta \;.
\end{equation}
COBE observed a CMB dipole amplitude of $\delta T \sim 3.3 (\pm 0.2)$ mK 
(which corresponds to a fluctuation 
$\delta T/T = 1.2 (\pm 0.03) \times 10^{-3}$). The 
corresponding velocity of Earth is:
$$ \vec{v}_\mathrm{\odot}-\vec{v}_\mathrm{CMB} \approx 365 (\pm 18) \;{\rm
km/sec}$$
towards the galactic coordinates $(l, b) = (265^{\circ}, 48^{\circ})$
(see \cite{Smo91}).
This motion is due to the vectorial sum of the motion of the Earth
around the Sun, of the Sun within the Galaxy, of the Galaxy within the
Local Group and of the peculiar motion of the Local Group, due to the
gravitational effects of large-scale density fluctuations.
The motion of the Earth with respect to the LG centroid is:
$$ \vec{v}_\mathrm{\odot}-\vec{v}_\mathrm{LG} \approx 308 \;\; {\rm km/sec} $$
towards $(l, b) = (107^{\circ}, -7^{\circ})$ and thus 
we find the velocity of the LG centroid with respect to the CMB:
$$ \vec{v}_\mathrm{LG}-\vec{v}_\mathrm{CMB} \approx 620 \;{\rm
km/sec}$$ 
towards $(l, b) = (277^{\circ}, 30^{\circ})\;.$

The Local Group velocity was originally thought as the result
of the attraction of the Local Supercluster (Virgo). However, there
is a residual velocity of $\sim 400$ 
km/sec that must be due to gravitational forces acting on the LG from 
distances greater than the Local Supercluster's centre-of-mass ($c z
\sim 1100$ km/sec). 
Many earlier studies pointed towards
the {\em `Great Attractor'}, a mass concentration of 
$\sim 5 \times 10^{16}$ $M_{\odot}$ 
located at a distance of 42 $h^{-1}$ Mpc and at low Galactic latitudes, 
as being the sole cause of a relatively local coherent motion,
in which the Local Group partakes (cf. \cite{Lyn88}, \cite{LC88}). 
Later studies, indicated that another very
massive and more distant ($\sim 140$ $h^{-1}$ Mpc)
attractor could play a significant role in shaping the local
dynamics (\cite{Sc89}, \cite{SVZ91}, \cite{PV91}). 
It seems that the coherence
scale of the velocity field could extend to even larger distances than
what originally thought 
(cf. \cite{BP98}, however for a different view see
\cite{D99}).

\subsection{Cosmological Parameters}
Based on (\ref{eq:ein1}) 
we can define some very important parameters like the {\em Critical
Density}, which is the density necessary to obtain
a flat Universe ($\Lambda=0$): 
\begin{equation}\label{eq:crden}
\rho_\mathrm{cr} = \frac{3 H_\mathrm{\circ}^{2}}{8 \pi G} 
= 1.88 \times 10^{-29} h^{2} \; \mbox{\rm gm cm$^{-3}$} 
\end{equation}
and the {\em Cosmological density parameter} $\Omega$, which is a 
unit-less measure of the density of the  Universe:
\begin{equation}\label{eq:omega}
 \Omega = \frac{\rho}{\rho_\mathrm{cr}}
\end{equation}
Furthermore, the constant of proportionality in Hubble's law, the {\em Hubble
constant}, is:
\begin{equation}
H_\mathrm{\circ} = 100 \; h \; \frac{\rm km}{\rm sec \; Mpc} 
= 1.023 \times 10^{-10} \;h \;\mbox{years}^{-1}
\end{equation}
Note that the necessity of parametrizing with $h$ was 
due to earlier discordant determinations of $H_{\circ}$. 
Today most studies converge to a value of $\sim 0.7$ (see section 2.4).

A convenient representation of these interrelations can be produced by
re-writing Friedmann's equation, in the matter dominated era 
(using \ref{eq:rho1}), as following:
\begin{equation}\label{eq:new_fre}
\frac{\dot{R}}{R} = H_\mathrm{\circ} \left(\Omega_\mathrm{m} (1+z)^3 +
\Omega_\mathrm{k} 
(1+z)^2 + \Omega_\mathrm{\Lambda} \right)^{1/2} \Longrightarrow 
H(z)= H_{\circ} E(z)
\end{equation}
were the contribution to the total density parameter from the curvature
and $\Lambda$ terms is:
\begin{equation}\label{eq:defin}
\Omega_\mathrm{k}=
-\frac{kc^2}{H_\mathrm{\circ}^2 R_\mathrm{\circ}^2} \;, \;\;\;\;\;
\Omega_\mathrm{\Lambda}=\frac{{\Lambda} c^2}{3 H_\mathrm{\circ}^2} \;.
\end{equation}
Note that $H(z)$ is called {\em Hubble function}.
It is evident that at the present epoch we obtain from
(\ref{eq:new_fre}) that $E(0)=1$ and thus:
\begin{equation}\label{eq:totOmega}
\Omega_\mathrm{m} +\Omega_\mathrm{k} +\Omega_\mathrm{\Lambda} =1
\end{equation}
which also holds for any epoch (evaluated directly from \ref{eq:ein1}).
Note that we can have a flat Universe ($\Omega_\mathrm{k}=0$) while having 
$\Omega_\mathrm{m}<1$ (as suggested by many different observations). 

\subsubsection{The Age of the Universe:}
Using (\ref{eq:new_fre}), evaluated at the present epoch, we have
$\dot{R}/R_{\circ} = H_{\circ} E(z)/(1+z)$
and from ${\rm d}R/R_{\circ} = -{\rm d}z/(1+z)^2$
we obtain the age of the Universe:
\begin{equation}\label{eq:age0}
t_{\circ} = \frac{1}{H_{\circ}} \int_0^\infty \frac{{\rm d}z}{(1+z) E(z)}
\end{equation}
For example, in an {\em Einstein-de Sitter} universe 
($\Omega_\mathrm{\Lambda}=\Omega_\mathrm{k}=0$) we have:
\begin{equation}\label{eq:age}
t_\mathrm{\circ} = \frac{2}{3 H_\mathrm{\circ}}
\end{equation}
while for a $\Omega_\mathrm{\Lambda} >0$ model we obtain:
\begin{equation}\label{eq:lam}
t^\mathrm{\Lambda}_\mathrm{\circ} = 
\frac{2}{3 H_{\circ}} \frac{1}{\sqrt{\Omega_\mathrm{\Lambda}} } \sinh^{-1} 
\left[\sqrt{\frac{\Omega_\mathrm{\Lambda}}{\Omega_\mathrm{m}}} \right]
\end{equation}
We therefore see that if $\Omega_\mathrm{\Lambda}>0$ we have that the age of
the Universe is larger than what is predicted in an Einstein-de Sitter
Universe.

\subsubsection{The $\Lambda \neq 0$ Universe:}
Due to the recent exciting observational indications for a positive
cosmological constant and the important consequences that this has to
our understanding of the Cosmos, we will briefly present this model.

Originally, the cosmological $\Lambda$-parameter was introduced {\em
ad hoc} by Einstein in his field equations in order to get a static 
solution ($\dot{R} = \ddot{R} = 0$). From (\ref{eq:ein2}) he
derived $R=(k/\Lambda_\mathrm{c})^{1/2}$ and inserting this into
(\ref{eq:ein1}) he obtained, in a matter
dominated ($p=0$) Universe:
\begin{equation}\label{eq:ein_sol}
\rho = \frac{\Lambda_\mathrm{c} c^2}{4 \pi G}
\end{equation}
where $\Lambda_\mathrm{c}$ is the critical value of $\Lambda$ for which 
$\dot{R} = \ddot{R} = 0$.
However, it was found that his solution was unstable and that 
small perturbations of the value of $\Lambda_\mathrm{c}$ 
would change drastically the 
behaviour of $R$. From (\ref{eq:ein1}) we see that if 
$k\le 0$, then $\dot{R}^2$ is always nonnegative 
for $\Lambda>0$, and thus the universe expands for ever, while
if $\Lambda<0$ then the universe can expand and then recontract again
(as in the $k=1$, $\Lambda=0$ case).

The recent SNIa observations (see section 3.2) and 
the CMB power-spectrum results (see section 3.1) have shown
that the {\em Standard} Cosmological paradigm
should be considered that of a flat, $\Omega_\mathrm{\Lambda}\simeq
0.7$, $\Omega_\mathrm{m}\simeq 0.3$ model. 
Thus we will consider such a model in the following discussion.
Evaluating (\ref{eq:ein1}) at the
present epoch, changing successively variables: 
$x = R^{3/2}$, $y= x (\Omega_\mathrm{m}/\Omega_\mathrm{\Lambda})^{1/2} \;
R_{\circ}^{-3/2}$ and $\theta = \sinh^{-1} y$
and then integrating, we obtain:
\begin{equation}\label{eq:lam0}
t = \frac{2}{3 H_{\circ} \sqrt{\Omega_\mathrm{\Lambda}}} 
\sinh^{-1} \left[ \left( \frac{\Omega_\mathrm{\Lambda}}{\Omega_\mathrm{m}}
\right)^{\frac{1}{2}} \left(\frac{R}{R_{\circ}}\right)^{\frac{3}{2}} \right]
\end{equation}
and 
\begin{equation}\label{eq:lam1}
R = R_{\circ} \left(\frac{\Omega_\mathrm{m}}
{\Omega_\mathrm{\Lambda}}\right)^{\frac{1}{3}}  
\sinh^{\frac{2}{3}}\left(\frac{3 H_{\circ}\sqrt{\Omega_\mathrm{\Lambda}}}{2} 
\; t \right) 
\end{equation}

\begin{figure}[t]
\mbox{\epsfxsize=12cm \epsffile{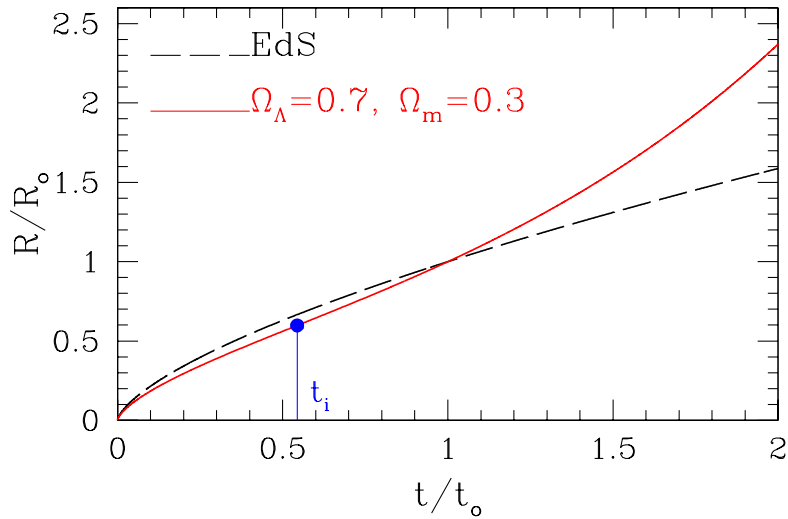}}
\caption{The expansion of the Universe in an Einstein de-Sitter (EdS) and
in the {\em preferred} $\Lambda$ model. 
We indicate the inflection point beyond which the 
expansion accelerates. It is evident that in this model we live in the
accelerated regime and thus the age of the Universe is larger than the
Hubble time ($H_{\circ}^{-1}$).}
\end{figure}
It is interesting to note that in this model there is an epoch which 
corresponds to a value of $R = R_\mathrm{I}$, where
the expansion slows down and remains in a quasi-stationary phase for some 
time, expanding with $\ddot{R} > 0$ thereafter (see Fig.1). 
At the quasi-stationary epoch,
called the inflection point, we have $\ddot{R} = 0$ and thus from 
(\ref{eq:ein1}) by differentiation we have:
\begin{equation}\label{eq:infl}
R_\mathrm{I} = 
\left(\frac{\Omega_m}{2 \; \Omega_\mathrm{\Lambda}} \right)^{\frac{1}{3}} 
R_{\circ}
\end{equation}
Now from (\ref{eq:lam0}) and (\ref{eq:infl}) we have that
the age of the universe at the inflection point is:
\begin{equation}\label{eq:tinfl}
t_\mathrm{I} = \frac{2}{3 H_{\circ}\sqrt{\Omega_\mathrm{\Lambda}}} 
\sinh^{-1}\left(\sqrt{\frac{1}{2}}\right) \;.
\end{equation}
The Hubble function at $t_\mathrm{I}$ is:
$$H(t_\mathrm{I})=H_{\circ} \sqrt{\Omega_\mathrm{\Lambda}} 
\coth \left(\frac{3 H_{\circ}\sqrt{\Omega_\mathrm{\Lambda}}}{2} \;
t_\mathrm{I} \right) \Longrightarrow 
H_\mathrm{I} = H_{\circ} \sqrt{3 \Omega_\mathrm{\Lambda}}$$
so if $t_{\circ} > t_\mathrm{I}$ we must have $H_{\circ} < H_\mathrm{I}$.

This is an important result 
because it indicates that introducing an $\Omega_\mathrm{\Lambda}$-term, and
if we live at a time that fulfils the condition $t_{\circ} >
t_\mathrm{I}$, we can
increase the age of the universe to comfortably fit the globular cluster 
ages while keeping the value of $\Omega_\mathrm{m} < 1$ and also a flat
($\Omega_\mathrm{k}=0$) space geometry. From (\ref{eq:tinfl}) and
(\ref{eq:lam}) and for the preferred values $\Omega_\mathrm{\Lambda}=0.7$
and $\Omega_\mathrm{m}=0.3$ 
we indeed obtain $t_{\circ}/t_\mathrm{I}\simeq 1.84$ 
(see also Fig.1), which
implies that we live in the accelerated phase of the Universe. Note
that in order for the present time ($t_{\circ}$) to be in the 
accelerated phase of the expansion we must have:
$\Omega_\mathrm{\Lambda}> 1/3$.

\subsubsection{Importance of $k$ and $\Lambda$ terms in global dynamics:}
Due to the recent interest in the $\Lambda > 0$, $k=0$ Universes, it is
important to investigate the dynamical effects that this term may have
in the evolution of the Universe and thus also in the structure
formation processes (see Fig.2).
We realize these effects by inspecting the magnitudes of the two terms in
the right hand side of (\ref{eq:ein1}). We have the density term:
$$ \frac{8 \pi G \rho}{3} = \frac{8 \pi G \rho_{\circ}}{3} (1+z)^{3} =
H_{\circ}^{2} \Omega_\mathrm{m} \left(1+z\right)^{3}$$ and from 
(\ref{eq:totOmega}) we have
\begin{equation}\label{eq:attr}
\frac{\Lambda c^2}{3} = H^{2}_{\circ} (1-\Omega_\mathrm{m})
\end{equation}
By equating the above two terms we can find the redshift at which 
they have equal contributions to the dynamics of the
Universe. Evidently this happens only in the very recent past:
\begin{equation}
z_\mathrm{c}=
\left(\frac{\Omega_\mathrm{\Lambda}}{\Omega_\mathrm{m}}\right)^{1/3} -1
\end{equation}
Observations suggest that $\Omega_\mathrm{m} \simeq 0.3$ and
$\Omega_\mathrm{\Lambda}\simeq 0.7$, and therefore we have 
$z_\mathrm{c}\simeq 0.3$,
which implies that the present dynamics of the universe are dominated by
the $\Lambda$-term, although for the largest part of the history 
of the Universe the
determining factor in shaping its dynamical evolution is the matter
content. 
\begin{figure}[t]
\centering{
\mbox{\epsfxsize=12cm \epsffile{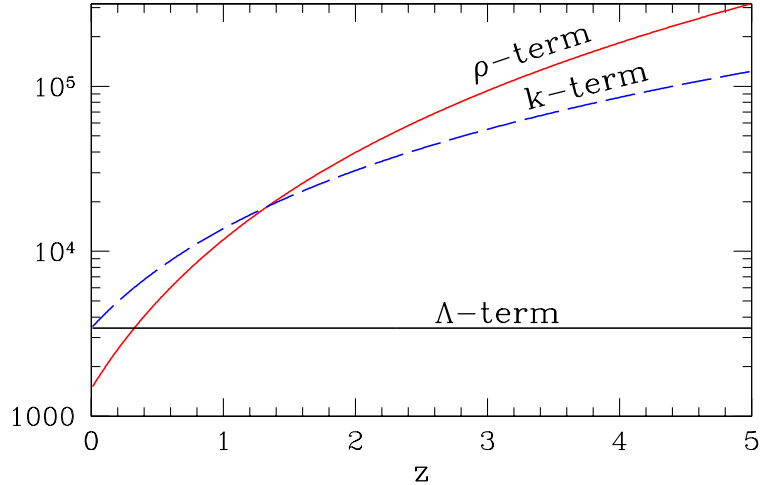}}
\caption{The strength of the three factors shaping the recent
dynamics of the Universe. Compare the strength of the $\rho$ and
$\Lambda$ term ($k=0$) and of the $\rho$ and $k$ term. We have assumed
$H_{\circ}=72$ km s$^{-1}$ Mpc$^{-1}$, $\Omega_\mathrm{m}=0.3$ and
$\Omega_\mathrm{\Lambda}=0.7$.}}
\end{figure}
Similar results are found for the $k$-term in
$\Lambda=0$ models. In this case we have from (\ref{eq:attr}) that
$$\frac{k c^2}{R^2} = H_{\circ}^{2} (1+z)^{2} |1-\Omega_\mathrm{m}| $$
and thus the redshift at which the density and curvature terms have
equal impact in the global dynamics, is:
$$ 
z_\mathrm{c} = \frac{1}{\Omega_\mathrm{m}}-2 \;.
$$
We see that as $z$ increases the density term grows faster than the curvature
term which is important only very recently. A similar line of argument
shows that also in the radiation dominated era the $\Lambda$ and $k$ terms 
do not affect the dynamics of the Universe.

\subsubsection{Density parameter as a function of ${\bf z}$:}
From (\ref{eq:new_fre}) and eliminating the curvature term, using
$\Omega_\mathrm{k}=1-\Omega_\mathrm{m}-\Omega_\mathrm{\Lambda}$, we
obtain the time evolution of the
density parameter $\Omega_\mathrm{m}$:
\begin{equation}
H^2=H_{\circ}^{2} \left[\Omega_\mathrm{m} (1+z)^3 +
\left(1-\Omega_\mathrm{m} - \Omega_\mathrm{\Lambda} \right)(1+z)^2 + 
\Omega_\mathrm{\Lambda}
\right] 
\end{equation}
and using (\ref{eq:crden}) and (\ref{eq:rho1}) we have:
\begin{equation}\label{eq:omeg_z}
\Omega(z) = \frac{\Omega_\mathrm{m} + \Omega_\mathrm{\Lambda}/(1+z)^{3}}{\Omega_\mathrm{m} + 
(1-\Omega_\mathrm{m})(1+z)^{-1} +
\Omega_\mathrm{\Lambda}(1+z)^{-1} \left[1/(1+z)^2 -1\right]}
\end{equation}
It is easy to see that whatever the value of $\Omega_\mathrm{m}$, 
at large $z$ we always have $\Omega(z)=1$. 

\subsection{Distribution of Matter in the Universe}
The task of determining and quantifying the 3-dimensional distribution of 
extragalactic objects, is not a trivial. Limitations are imposed on us
by our position in space-time, and also the fact that we are located in the
plane of a dusty spiral galaxy, which means that 
the light-absorbing interstellar dust will cause the magnitudes and 
sizes of galaxies to decrease, especially if we observe towards the
Galactic plane.

Generally in order to be able to quantify the geometry and topology of 
the large-scale structure of the Universe and to discriminate between the 
competing scenarios of structure formation, 
two at least issues should be addressed:

\noindent
{\bf (a)} The quantification, in an objective manner, of the 
observed structure on large scales. 
For this we need to observe the positions and redshifts of galaxies
tracing large-volumes, large enough to be considered a fair sample of the 
Universe. If not 
we can be influenced by local {\em anomalies} (like a local attractor
or a local underdense region) and thus
the interpretation of our results could lead us to erroneous
conclusions. Today there are large samples available, complete to some 
apparent magnitude limit which are deep enough for volume-limited samples 
to be extracted from them (for example the {\sc pscz, ors, las
campanas, apm, ssrs}, 2dF, {\sc sdss} catalogues). 

\noindent
{\bf (b)} The application of well defined and easily applicable 
statistics, 
which are able to distinguish between the different theoretical
scenarios of structure formation and 
to compare between these scenarios and the observed topology and
geometry of the large-scale structure of the Universe.

Of historical interest are the first attempts to quantify the
distribution of galaxies by Bok and Mowbray (in the mid 1930s). 
They compared the variance of galaxy counts $N$ with that expected from a 
Poisson distribution (ie, compared $\langle N
\rangle^{1/2}$ with $ \sigma \equiv [\frac{1}{M} \sum (N_\mathrm{i} 
- \langle N 
\rangle)^{2}]^{1/2}$, where $M$ is the number of fields) and concluded
that $\sigma \gg \langle N \rangle^{1/2}$ an indication that {\em galaxies
cluster}. More sophisticated statistics appear in the mid-50's when
the large Lick galaxy counting project started giving its first
results (cf. \cite{Pee80}).
Through the years a wide variety
of statistical techniques have been developed in
order to
study the geometry and topology of the distribution of matter on
large-scales (for reviews see \cite{Borg95} \cite{Coles_rev}).

\subsubsection{Galaxies:}
Galaxies are the basic units of the distribution of matter in the
universe. They are conglomerations of stars, dust and gas 
having typical sizes of a few kpc. Only recently, in the 1920s,
 was it realized that the fuzzy nebulae, that we now know to be
galaxies, were not part of the Milky-Way.
There appear to be many different types of galaxies, having different
shapes, amounts of gas and dust, as well as star-forming regions. Many
attempts have been made to produce a reliable classification
scheme that bears relation to physical quantities.
Galaxies come into three main categories: ellipticals, spirals and irregulars, 
in which the basic difference, as their names imply, is their apparent 
shape. 
However, further subdivision is essential in order to classify
reliably the large variety of galaxies. 

The first attempt to classify galaxies was that of Hubble in 1926.
His principal morphological classification was provided
by the ``tuning-fork'' diagram, in which the
position of a galaxy was determined by the size of its nucleus and on
the spiral arm tilt. For historical reasons, galaxies at the left of the
diagram are called ``early types'' while at the right ``late types''.
Although this diagram was produced on the basis of appearance there seem
to be a link between morphology and galaxy evolution since the
integrated colours and spectral types monotonically increase from left
to right (\cite{Berg98} and references therein).
The fraction of the different
galaxy types is: $\sim 10\%$ ellipticals (E), 
$\sim 20\%$ SOs, $\sim 65\%$ spirals (S) and $\sim 5\%$ irregulars (Irr).

The spiral galaxies consist of three 
components: disk, bulge and halo. The disk is very thin (a few 
hundred pc thick). The surface brightness of stars
in a typical disk falls exponentially with radius:
\begin{equation}
I_\mathrm{s} = I_{\circ} \exp{[-r/b]}
\end{equation} 
where $b$ is the scale-length of the disk (typically $\sim 4$ kpc).
The spiral structure is thought to be the result of rotating density waves
which produce shocks in the gas-rich disk which leads to star formation.

The elliptical galaxies however, are 
ellipsoidal systems with an old population of stars and very little gas. They
have a {\em de Vaucouleurs} luminosity distribution:
\begin{equation}\label{eq:dev}
I_\mathrm{s} = I_\mathrm{e} \exp{[-7.67((r/r_\mathrm{e})^{0.25} - 1)]}
\end{equation}
where $r_\mathrm{e}$ is the {\em effective} 
radius enclosing half of the total light.
Therefore $I(r)$ falls off more slowly than $r^{-2}$ for
$r<r_\mathrm{e}$ and more
rapidly for $r>r_\mathrm{e}$ 
(this formula also fits the bulges of SOs and spirals).
The mass of elliptical galaxies can vary widely; from dwarfs with $M \sim 
10^{7}$ $M_{\odot}$ to supergiants with $M \sim  10^{12}$ $M_{\odot}$ 

\subsubsection{Groups \& Clusters of Galaxies:}
Galaxy clusters occupy a special position in the hierarchy of cosmic 
structures in many respects. Being the largest bound structures in the
universe, they contain hundreds of galaxies, large amounts of Dark
Matter (DM) and hot X-ray
emitting gas, and thus can be detected at large redshifts. Therefore, they
appear to be ideal tools for studying large-scale structure, 
testing theories of structure formation and extracting invaluable 
cosmological information  
(see recent reviews \cite{BG00}, \cite{Bac_r00})
Groups of galaxies are typically systems that contain $\lesssim 10 - 20$ 
galaxies (for a recent review see \cite{group}).

There are several classification schemes for clusters. A simple and clear-cut
one is between {\em regular} and {\em irregular} clusters.
\begin{itemize}
\item[$\otimes$] {\bf Regulars:}
They have smooth and symmetric structure, with high central density ($\ge 
10^{3}$/Mpc$^{3}$), a small fraction of spiral galaxies ($\lesssim 20\%$), 
high velocity dispersion ($\sim 10^{3}$ km/sec) and high X-ray luminosity.
The high velocity dispersion as well as the smooth structure is
considered to be
evidence that these clusters are in virial equilibrium (relaxed). About
$50\%$ of all clusters are regulars.

\item[$\otimes$] {\bf Irregulars:}
They have lumpy structure with evident substructures, 
lower velocity dispersion and lower X-ray 
luminosity. The fraction of spirals is higher than that in regular clusters 
($\gtrsim 40\%$).
\end{itemize}
Another distinct class of clusters is those having a central very bright 
galaxy (BCG or cD's). 
The BCG galaxies are giant ellipticals, some with multiple nuclei (cD's)
and masses $M\gtrsim 10^{12} \; M_{\odot}$. There are different views
regarding their formation history; one advocates
that they form by ``galactic cannibalism" where dynamical 
friction causes cluster galaxies to spiral towards the cluster core and then
strong tidal effects take place (cf. \cite{Merr}, \cite{Bar01}),
the other
advocates that they form at {\em special} locations, which are the
kernels around which clusters will eventually form by anisotropic accretion
of matter (cf. \cite{We94} and references therein). 
Furthermore:
\begin{itemize}
\item[$\blacklozenge$] There is a well defined relationship between galaxy 
density and fraction of Hubble types. The fraction of E's and SO's 
increases smoothly with increasing galaxy-density (\cite{Dr80}).
An interesting question is whether the above relation is a direct 
result of the galaxy formation processes, or an evolutionary effect.
\item[$\blacklozenge$] Clusters are highly flattened systems even more 
than elliptical galaxies. This fact is not due to rotation 
(cf. \cite{Dr81}, \cite{Rood72}), and therefore it either 
reflects the initial conditions of their formation and/or the
tidal effects acting between proto-clusters. In fact, the high peaks of an 
initial Gaussian density field, the possible sites of 
cluster formation, are non-spherical (\cite{Bar86}). 
The cluster mean projected ellipticity 
is $\varepsilon \sim 0.5$ and there is evidence that clusters are more
prolate-like (cf. \cite{BPM00}).
\item[$\blacklozenge$] Neighbouring clusters tend to 
point to each other up to distances of $\sim 20 - 30$ $h^{-1}$ Mpc 
(cf. \cite{Bin82}).
Dynamically young (irregular) 
clusters show a tendency to be more aligned with their neighbours and are
preferentially found in high-density environments (\cite{PB02}, 
\cite{Schuk}).
\end{itemize}

For large-scale clustering, dynamical and cosmographical studies, 
it is extremely important to compile large, whole-sky catalogues 
of groups and clusters. One of the first and extensively studied
such catalogue is the {\sc abell/aco} catalogue \cite{ab1},
which was based on an eyeball selection procedure from sky survey
plates. This all-sky sample contains 4073 rich clusters,
nominally complete to a redshift $z=0.2$. Its obvious limitations, due
to the eyeball selection procedure, were 
superseded by the objectively selected {\sc apm} clusters (\cite{Dalt}) which
were based on the {\sc apm} galaxy catalogue
containing $4\times 10^6$ galaxies (\cite{Mad}). 
This cluster catalogue covers latitudes $b<-35^{\circ}$
and contains 950 clusters, typically poorer than the {\sc abell/aco} one's.
Furthermore, with the {\sc Rosat} whole-sky X-ray survey it was possible to
construct X-ray selected cluster catalogues (cf. \cite{Ebel}), 
which suffer less from
projection effects, which can produce {\em phantom} clusters. 
For example the {\sc Reflex} X-ray cluster sample contains
$\sim 450$ clusters (\cite{reflex}).

\begin{figure}
\centering{
}
\caption{
(Included separately as a JPG image)
An example of a filamentary supercluster: The 2D distribution of {\sc apm}
clusters members. The inserted plots show
the smooth 2D galaxy density map of some of the cluster members. For
the case of A3128 we overlay the X-ray emission contour plots. 
The cluster's elongation along the filamentary
supercluster is evident.}
\end{figure}

\subsubsection{Superclusters, Filaments \& Voids:}
Superclusters are aggregates of clusters, groups and galaxies. They
are the largest, isolated, but dynamically un-relaxed due to their size, 
objects in the large scale distribution of matter and thus 
they are ideal probes of the initial conditions that gave rise to
them. This fact is because typical peculiar
velocities of clusters are $v_\mathrm{pec} \sim 10^{3}$ km/sec and therefore
in a Hubble time ($1/H_{\circ}$) they can move no more than $\sim$10 
$h^{-1}$ Mpc, which is substantially smaller than the scale of a typical
supercluster. 
Superclusters can be identified in 3-D catalogues of clusters but also 
in 2-D projections of galaxy distributions. Regions of
high density in clusters can be identified also on the {\sc apm} galaxy map.

The large scale 
clustering pattern of galaxies is expected to 
be characterized by a filamentary 
and sheet-like distribution (cf. \cite{Zeld}, \cite{Ein01}, \cite{Weyg}). 
Indeed many authors have
been finding that the vast majority of the superclusters are flattened 
with a predominance of filament-like shapes (cf. \cite{BPR01} and
references therein). Figure 3 shows the 2-dimensional
projection of a filamentary supercluster containing 5 {\sc apm} clusters. 


Superclusters are not centrally condensed objects (like clusters) and 
their typical size is $\sim 30 - 50 \; h^{-1}$ Mpc. 
However larger structures, with a length $\sim 200 \; h^{-1}$ Mpc, 
may exist (cf. \cite{Tul} and references therein).
Detailed studies have shown that elongated bridges ($\sim 30$ $h^{-1}$
Mpc) of galaxies have been found to connect rich clusters.
Since wide-angle three-dimensional surveys became available, the filamentary 
distribution of galaxies has been a constantly observed feature. 
Even the original 
{\sc Cfa} survey (cf. \cite{Huc}) showed networks of 
filaments mostly connecting rich clusters of galaxies and large
voids (cf. \cite{Pee01}), a fact which has
been confirmed by all the recent surveys ({\sc ssrs, ors, pscz} etc).
Voids are regions of density well below the average value. In all 
deep  radial-velocity surveys, the velocity distribution shows striking empty 
regions were no (or very few) galaxies exist.
It is an extremely difficult task to identify voids in 2-D projections of 
galaxy distributions since the projection will tend to smooth-out any such 
structure. An
extremely interesting 
question, relevant to theories of structure formation, is whether the voids 
are empty of luminous matter or empty of all matter.

\newpage
\section{Distance Scale, Hubble Constant \& the Age of the Universe}
One of the most important parameters in
determining the fate of the Universe as a whole, is the present day
expansion rate of the Universe, which is encapsulated in the value of
the Hubble constant $H_{\circ}$. Its value sets the age of 
the Universe and specifies the value of the
critical density $\rho_\mathrm{cr}$, and through this
route the geometry of the Universe.

From the Hubble law (\ref{eq:hub}) 
it is evident that in order to determine its
value we need to determine the expansion velocity
(redshift) as well as the distance of
extragalactic objects, but within a redshift such that space-curvature
effects do not affect distances (\ref{eq:lumd}).
A further concern is that
local gravitational effects produce {\em peculiar}
velocities, that are superimposed on the general expansion. This can
easily be seen in the toy model of (\ref{eq:hub}), if we allow $\chi\equiv
\chi(t)$. Then
$$v=H_{\circ} d + R \dot{\chi} \;,$$
with the factor on the right being the peculiar velocity. Since the
observer as well as the extragalactic object, of which the distance we
want to measure, have peculiar velocities then the above equation
becomes:
\begin{equation}\label{eq:hub1}
c z  = H_{\circ} d + \left( \vec{v}(d) - \vec{v}(0) \right) \cdot \hat{r}
\end{equation}
where $v(0)$ is the velocity of the observer and $\hat{r}$ is the
unit vector along the line-of-sight to the extragalactic object. It is
then obvious that in order to measure $H_{\circ}$, 
the local velocity field should be measured and the
extragalactic distances corrected accordingly.
It is easily seen that if both observer and galaxy take
part in a coherent bulk flow, having the same amplitude at the
observer and galaxy positions, then the right-hand part of
(\ref{eq:hub1}) vanishes.
In general however, one needs good knowledge of the velocity field in order
to correct distances adequately.

\subsection{Distances of Extragalactic Objects}

Our only means of obtaining information and therefore knowledge of the 
structure and dynamics of the Universe (on all different scales) are
through the electromagnetic radiation that we receive. 
Therefore it is of primary importance to define a system
of measuring luminosities taking also into account that the Universe expands 
and that light loses energy through a variety of processes.

If we assume that light propagates with no loss of energy, then the apparent
luminosity of a source $l$, is related to its absolute luminosity $L$, by:
\begin{equation}\label{eq:1}
l = \frac{L}{4\pi r^{2}}
\end{equation}
where $r$ is the distance to the source. We can see the extreme
importance of determining the pair (${l,L}$), since such knowledge
would provide the distance of the source, $r$.

Due to historical mostly reasons we use a logarithmic 
brightness system by which an object with a 
magnitude of 1 is 100 times brighter than an object 
with a magnitude 6. We have:
\begin{equation}
m = -2.5 \log_{10} l + c_{1} \;\;\;\;\;\;\;\; M = -2.5 \log_{10} L + 
c_{2}
\end{equation}
where $m$ is the apparent magnitude and $M$ the absolute one. Therefore, using
(\ref{eq:1}) we have:
\begin{equation}\label{eq:2}
m - M = 5 \log_{10} r + c_{3}
\end{equation}
where $c_{1,2}$ are constants which depend on the filter used and $c_{3}$ is 
that value for which $m=M$ at a distance of 10
parsecs (see section 2.3) from the Earth, and thus $c_{3} = -5$.
In extragalactic astronomy, instead of $pc$, we use $Mpc$ and therefore
(\ref{eq:2}) becomes:
\begin{equation}\label{eq:mM0}
m - M = 5 \log_{10} r + 25
\end{equation}

The above definitions are somewhat $ideal$, since in the real world we do not 
observe the total apparent magnitude, but that corresponding to the 
particular range of spectral frequencies, that our detector is sensitive to, 
and those allowed to pass through Earth's atmosphere. If we detect
$l_\mathrm{\nu}$ over a range of frequencies $(\nu \pm \delta\nu)$, then
the observed apparent magnitude is
$ m = -2.5 \log_{10} \int_\mathrm{\delta\nu} l_{\nu}
\D\nu + c$. However, 
neither the atmosphere nor the detectors have a sharp $\nu$ 
limit and therefore it is better to model
these effects by a sensitivity mask $F_\mathrm{\nu}$, 
and the observed apparent magnitude is then:
\begin{equation}\label{eq:Fv}
m_\mathrm{F_\mathrm{\nu}} = -2.5 \log_{10} \left[ \int_{0}^{\infty} 
F_{\nu} l_{\nu} \D\nu \right]  + c 
\end{equation}
If $F_\mathrm{\nu} = 1$ 
then the apparent (or absolute) magnitude of a source is
called the $bolometric$ magnitude.

How do the above definitions change by taking into account the fact that 
the Universe expands? To answer this we need a metric of space-time,
which in our case is the  {\em Robertson-Walker} metric.
Since light travels along null-geodesics, a fundamental
concept of distance can be defined
by the corresponding light-travel time, which is called
{\em proper distance}.
If a light signal is emitted at a galaxy  ${\cal
G}_{1}$ from the
coordinate position $(r_{1}, \theta_{0}, \phi_{0})$ at time $t=0$
and received by an observer at ${\cal G}_{0}$ 
at $(r_{0}, \theta_{0}, \phi_{0})$, then
these events are connected only by the light signal and since all 
observers must measure the same speed of light, it defines a very fundamental
concept of distance. Obviously, it depends on the curvature of space,
and since $\D s=0$ we have from (\ref{eq:rwm}):
\begin{equation}
 d_\mathrm{pro}(t) = R(t) \int_{0}^{r_{1}} \frac{dr}{\sqrt{1 - k r^{2}}} =
   \left\{ \begin{array}{lll}
R(t) \; \sin^{-1}r_{1} & \mbox{$\;\;\;\; k = +1$} \\
R(t) \; r_{1} & \mbox{$\;\;\;\; k = 0$} \\
R(t) \; \sinh^{-1}r_{1} & \mbox{$\;\;\;\; k = -1$} 
 \end{array}
\right\} 
\end{equation}
In the expanding Universe framework, the expressions 
(\ref{eq:1}) and (\ref{eq:2}) change, to: 
\begin{equation}
l = \frac{L}{4\pi \; d^2_\mathrm{pro} 
(1 + z)^{2}} \;,\;\;\;\;\;
\;\;\;\;  m - M =5 \log_{10} d_\mathrm{L} + 25
\end{equation}
where $d_\mathrm{L} \equiv d_\mathrm{pro}(1+z)$ 
is the {\em luminosity distance}. It
is obvious that the distance measure of an extragalactic object
depends on the underlying Cosmology. A proper derivation of the
luminosity distance (cf. \cite{Weinb}, \cite{Peac00}) provides the following
expression:
\begin{equation}\label{eq:lumd}
d_\mathrm{L} = \frac{2 c}{H_{\circ} \Omega^{2}} \left[ \Omega z + 
(\Omega - 2) \left(\sqrt{1 + z \Omega} - 1 \right) \right]
\end{equation}
where $\Omega$ contains all the contributions (mass, energy
density, curvature). 

Another important distance definition is that of
the angular-diameter. It is based on the fact that a length $l$,
subtends a smaller angle, $\theta$, the further away it is ($l\propto
1/\theta$) and it is given from:
relation to:
\begin{equation}\label{eq:luma}
d_\mathrm{\theta} = \frac{d_\mathrm{L}}{(1+z)^2}
\end{equation}
Note that this notion of distance is used to
derive the CMB power spectrum predictions of the different
cosmological models (see section 3.1).

\subsection{Biases affecting Distance Determinations}
Many effects, related either to our position in space (being on the
plane of a dust-rich spiral galaxy), to natural limitations
(introduced for example by the expansion of the Universe) 
or to detector related issues,
introduce systematic biases that affect our ability to measure
accurately distances.
Below, I list a few of the most important such effects.

\subsubsection{K-correction:}
Since bolometric magnitudes are not possible to measure,
but rather magnitudes over a particular wavelength range,
it is important to correct these magnitudes
for the effect of the expansion of the Universe.
These considerations result in modifying the distance modulus by a
factor, the so-called $K-$correction factor:
\begin{equation}
 m - M = 5\log_{10} d_\mathrm{L} + 25 + K(z)
\end{equation}
This factor arises from the fact that when we measure the magnitude of 
sources at large distances and at a particular frequency, say 
$\nu_\mathrm{o}$, 
we receive light emitted from a different part of the spectrum,
$\nu_\mathrm{e}$. It could well 
be that in this latter part of the spectrum the extragalactic object is 
particularly fainter or brighter than in the nominal one, $\nu_\mathrm{o}$.
Furthermore, a combination of different factors; 
evolution, intervening absorption 
processes or detector sensitivity for example, 
result in energy losses as a function of 
wavelength, which can be expressed by a mask $F(\nu_\mathrm{o})$ 
(\ref{eq:Fv}). 
Knowing $F(\nu_\mathrm{o})$ one can estimate the $K$-factor by 
integrating the spectrum at the source rest frame. 
For example, such calculations have 
shown that a typical value for spiral galaxies at $z=1$ is $K\approx 2$ (in 
general $K(z) \propto z$). 

\subsubsection{Malmquist bias:}
Due to the nature of astronomical observation
there is a minimum flux density above which we select extragalactic
objects. As we look deeper in the Universe we tend to pick up a
relatively larger fraction of intrinsically bright objects (ie.,
only the brighter end of the luminosity function is sampled).
This bias arises when determining distances 
from apparent magnitude limited samples. 
If the individual absolute 
magnitudes $M_\mathrm{i}$ of a sample of extragalactic objects 
have a
Gaussian distribution around $\langle M \rangle$ with dispersion
$\sigma$, then
this bias, for the case where the distribution of extragalactic objects
is considered homogeneous, is given by:
\begin{equation}
\Delta(M) = 1.382 \sigma^2
\end{equation}
How does this bias affect the determination of extragalactic distances?
The inferred distances of extragalactic objects are typically smaller
than their true distances. From (\ref{eq:mM0}) we have that:
$$r_\mathrm{cor} \approx r_\mathrm{raw} \; 10^{1.382 \sigma^{2}/5}$$
\begin{figure}[t]
\centering{
\mbox{\epsfxsize=12cm \epsffile{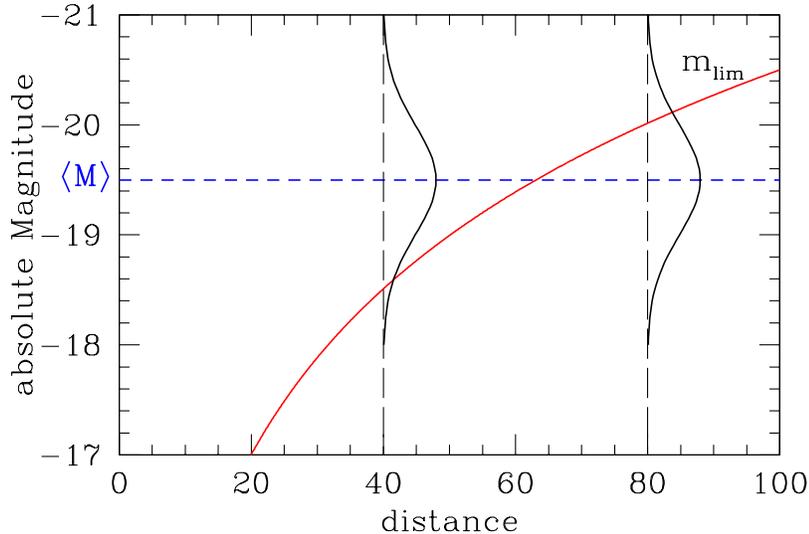}}
\caption{Illustration of Malmquist bias: Only objects with $M$ above
the $m_\mathrm{lim}$ limit can be observed. At different distances different
portions of the distribution around $\langle M \rangle$ can be observed.}}
\end{figure}
We illustrate this bias in Fig.4;
as the distance increases, $M(m_\mathrm{lim})$ becomes brighter and therefore 
the brighter end of the luminosity function is sampled.
For a larger $m_\mathrm{lim}$ (deeper sample) the value of 
$M(m_\mathrm{lim})$ 
increases (less luminous) so we have a smaller $\Delta(M)$. 
Conversely, for a given $m_\mathrm{lim}$ and $\langle M \rangle$, 
the bias increases with distance.

Note that we have considered a fairly straight-forward case, ie. that of a 
sample with a unique $\langle M \rangle$ value. In real samples of 
extragalactic 
objects we have a range of such values and therefore this bias is not easily
seen.

A related bias that also affects extragalactic distance determinations is 
the fact that there are larger numbers of objects at 
larger distances and therefore within a given range of estimated distances, 
more will be scattered by errors from larger to smaller distances 
than from smaller to larger ones.

\subsubsection{Galactic absorption:}
Interstellar gas and dust absorbs the background light with dust scattering
more efficiently the blue light, and thus the background light appears
artificially reddened. From simple geometrical considerations it is easy to
show that the flux $l_\mathrm{\nu}$ 
of an extragalactic source, transversing a
Galactic layer of thickness $\D s$, at an angle $b$ from the equatorial plane,
suffers losses $\delta l_\mathrm{\nu}/l_\mathrm{\nu} \propto \D s \;
\mbox{cosec}(b)$ and therefore:
\begin{equation}
\frac{\D l_\mathrm{\nu}}{\D s} = - 
\kappa_\mathrm{\nu} l_\mathrm{\nu} \mbox{cosec($b$)}
\end{equation}
where the constant of proportionality $\kappa_{\nu}$ is the absorption 
coefficient at the spectral frequency $\nu$. Therefore, integrating we have:
\begin{equation}
l_\mathrm{\nu} = l_\mathrm{\nu}^{o} \; 
\exp{\left[ - {\cal A} \; \mbox{cosec$|b|$} \right]}
\end{equation}
where the integration constant $l_\mathrm{\nu}^\mathrm{o}$ 
is the incident and $l_{\nu}$ is the observed flux,
 while ${\cal A} = \int \kappa_\mathrm{\nu} \D s$ is the {\em optical
thickness}. Therefore to take into account this effect 
(\ref{eq:1}) should change to:
\begin{equation}\label{eq:corrected}
l_\mathrm{true} = \frac{l_\mathrm{raw}}
{\exp\left[-{\cal A} \mbox{cosec}|b|\right]} = 
\left(\frac{L}{4\pi r^{2}} \right) \; \exp{\left[ {\cal A} \; 
\mbox{cosec$|b|$}\right]} \;.
\end{equation}
Values of ${\cal A}$ slightly vary for different spectral frequency bands, but
a generally accepted value, in V, is $\sim 0.2$. We see from 
(\ref{eq:corrected}) and (\ref{eq:mM0}) that 
\begin{equation}
 r_\mathrm{true} \approx r_\mathrm{raw} \exp{[-{\cal A} \; \mbox{cosec}|b|/2 ]}
\end{equation}
ie., the distance of an extragalactic source at a given absolute magnitude can
be significantly overestimated at low galactic latitudes if this effect is 
not taken into account. Note however that the cosec$(b)$ model is 
oversimplified since the distribution of gas and dust in the Galaxy is 
rather patchy. 

\subsubsection{Cosmological Evolution:}
As we look back in time we see a distribution of
extragalactic objects (normal galaxies, AGN's, clusters) 
in different evolutionary stages. 
It may well be that their luminosity and/or mean number density
is a function of cosmic time, a fact that will affect distance
determinations based on local calibrations of the relevant scaling-relations.
\subsubsection{Aperture effect:}
Since galaxies are extended objects with no sharp outer boundaries, their 
photometric measures will depend also on the size of the telescope
aperture since at different distances different fraction of a galaxy
will fit in the aperture.
This is a distance--dependent effect, since diameters scale like
$1/r$, and therefore it may affect the distance estimate.

\subsection{Distance Indicators:}
In order to develop the distance ladder from local to distant cosmic
objects, one starts from the local distance scale (for a detailed
account see \cite{RR85} \cite{RR88}).
\subsubsection{Galactic Distance Indicators:}
The primary method used to estimate the distances to nearby stars is that of
{\em Trigonometric Parallax}. As the Earth
orbits the Sun, we view the Universe from different points of the orbit
throughout the year. This difference is maximum  every 6 months when
the Earth is on the opposite side of its orbit around the Sun.
This kinematic method provides the basic unit of distance, the {\em parsec},
 used by astronomers. It is defined
as the distance at which a star would have a trigonometric parallax of 
1$^{''}$ (parsec = {\em par}allax + {\em sec}ond):
$$ 1 pc = 3.086 \times 10^{13} \;\; \mbox{km} \; = \; 3.26 \;\; \mbox{light years} 
$$
Note that this method is effective out to $\sim 60$ parsecs and that
the nearest star to us ($\alpha$-Centauri) is at a distance of 0.75
parsecs.

Among the many distance indicators, used to determine distances within our
Galaxy, a particularly interesting one
is the {\em Main sequence fitting} method. This takes advantage of the
fact that stars in globular clusters are at a common distance and that
there is a unique correlation between spectral stellar type 
and absolute luminosity (the {\em H-R} diagram). 
Therefore by measuring the distance, via a kinematic method, to one
relatively near globular cluster, one sets the zero-point of this
method and then by observing the apparent magnitude - spectral type
distribution of other globular clusters, one can determine their distance. 

\subsubsection{Extragalactic Distance Indicators:}
The next step is based on {\em Cepheid Variable Stars}. 
This method has been traditionally used 
within our Galaxy and in the nearby Large Magellanic Cloud (LMC), 
but with the {\em Hubble Space Telescope} 
it has been successfully used out to $\sim 20$
Mpc (cf. \cite{Freed01}).
A strong and tight relationship exists
between the intrinsic luminosity of these stars and their 
period of variation (pulsation) which results in a
Period-Luminosity relation:
\begin{equation}
L \propto \log P^{1.3} \Longrightarrow \log L = \mbox{\rm zero-point} + 1.3
\log P
\end{equation}
Once this relation has been calibrated, it provides the absolute
luminosity of the distant Cepheid stars and via (\ref{eq:mM0}), 
the distance of their host galaxy. Although this relation has a
scatter, in the $I-$band, of only $\pm 0.1 \; mag$, systematic effects
may exist. For example, a serious concern is whether there is any
environmental dependence of the relation. It has been suggested that
a different metalicity of the host galaxy may significantly affect the
zero-point of the relation and thus the determined distance. 
These effects can be taken into account and this
method has proved to be fundamental in the recent determinations of
$H_{\circ}$, because it provides the link between the primary galactic
indicators and the {\em local} extragalactic ones, which then provide the
calibration for other secondary indicators operating in much
larger distances (cf. \cite{Freed01} and references therein).

In developing the distance scale, we now need effective
indicators that can be used to very large distances.
Other scaling relations have been found between a distance dependent
(ex. brightness, diameter) and a distance independent (ex. rotational velocity,
stellar velocity dispersion) quantity. It is
evident that from such relations one can extract distance information.
The main assumption in such a use of these scaling 
relations is that they are not environment-dependent (which has been shown
to be a valid assumption).
Such a relation for spiral galaxies is the {\em Tully-Fisher relation}
\cite{TF}, which relates the 
rotational velocity of a spiral to its total infrared luminosity:
$ L_\mathrm{ir} \propto V^{4}_\mathrm{rot}$ or its total blue luminosity:
$ L_\mathrm{b} \propto V^{\alpha}_\mathrm{rot}$ 
with $\alpha \sim 2.4 - 2.8$. It
has a reasonable theoretical justification:
Rotational velocities in spirals are related to mass according to
\begin{equation}
V_\mathrm{rot}^2 \propto \frac{M}{R}
\end{equation}
Assuming that all spirals have the same surface brightness $S$, then
\begin{equation}
S \propto L/R^2 \Rightarrow L\propto R^2
\end{equation}
If the mass to light ratio is constant then
\begin{equation}
\frac{M}{L} \propto V_\mathrm{rot}^2 L^{1/2} 
\Rightarrow L\propto V_\mathrm{rot}^4
\end{equation}

For ellipticals a similar relation holds, the {\em
Faber-Jackson} relation \cite{FJ}, which relates the
absolute luminosity of the galaxy with the stellar velocity dispersion ($L
\propto \sigma^{3-4}$) or a variant, the so-called 
{\em $D_\mathrm{n}-\sigma$ relation} (cf. \cite{Dr}), which
relates the diameter $D_\mathrm{n}$ of an elliptical 
(defined as that within which the mean surface
brightness is 20.75 {\em mag arc sec$^{-2}$} in B) to the stellar velocity
dispersion $\sigma$: $ D_\mathrm{n} \propto \sigma^{x}$ 
with $x \sim 1.2 - 1.3$ . 
The typical accuracy of these distance estimations is $\sim 20\%$ and
the usual assumption is
that they do not evolve with redshift over the scales used, and that
systematic effects can be corrected for. However, there are
some indications for an evolution of
the $B$-band {\em Tully-Fisher} relation in
the interval studied $0.1<z<1$ \cite{Zie}.

Another very important distance indicator, which can be used to
large distances, is the {\em Supernova SNIa  brightness at maximum}.
It is thought that the maximum luminosity of such supernovae (explosion
of white dwarfs in binary systems which become gravitationally unstable
- reaching the Chandrasekhar limit - 
due to the accretion of matter from the secondary - see \cite{Liv}) 
is a Universal
constant and since the intrinsic luminosity of a SNIa is high,
they can be seen out to cosmological distances.
Furthermore, a correlation was found between the supernova
luminosity and the brightness decay time, which provides a further
parameter that reduces the scatter in luminosity to $\pm 0.3$ mags.
Using this distance indicator one can construct the Hubble diagram to
very large-distances and thus determine the deceleration of the
Universe by mapping the region of the Hubble diagram that
deviates from linearity. This has been recently achieved by two
independent groups, the {\em Supernova Cosmology Project - SCP}
\cite{Perl95} and the {\em High-z Supernova Search Team - HZT} 
\cite{Sc98},
which have found that the derived Hubble diagram is that expected from
a accelerating expansion, which can be provided by a non-zero
cosmological constant (see Fig.5).

Although I do not plan to present all the secondary distance
indicators, one that is potentially very important and susceptible to
small uncertainties and systematics, is the {\em surface brightness
fluctuation} method. This method is based on the fact that the discreteness of
stars within galaxies depends on distance. This methods 
has an accuracy of $\sim 5\%$ in distance.

\begin{figure}[t]
\hfill
\mbox{\epsfxsize=12cm \epsffile{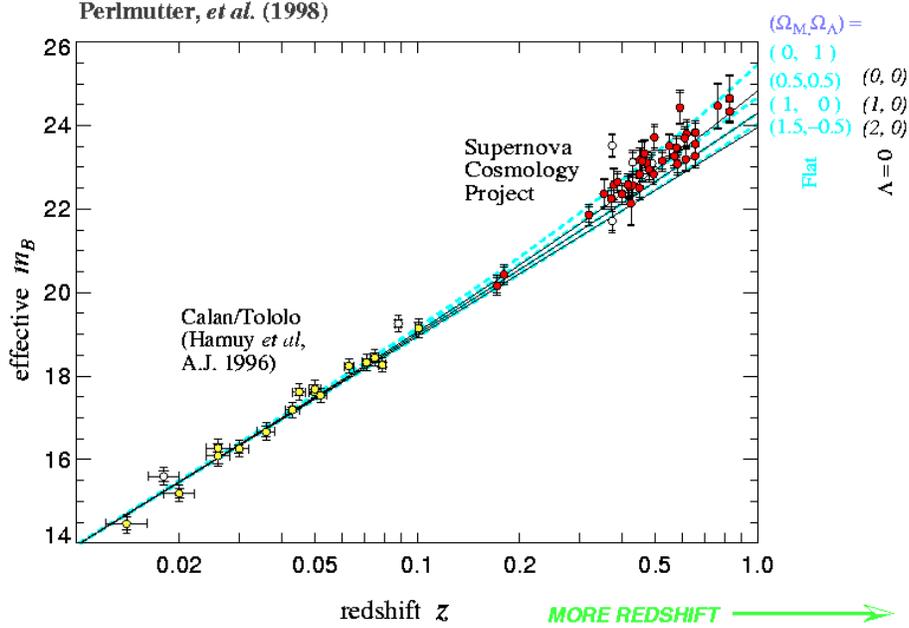}}
\caption{The SN Ia Hubble diagram {\em vs} Cosmological model
predictions (from \cite{Perl99} with permission).} 
\end{figure}

\subsubsection{Direct Distance Indicators:}
Clusters of galaxies are filled with hot and transparent gas which can
be fitted by a thermal Bremsstrahlung 
spectrum with $T\sim 5 \times 10^{7}  - 10^{8}\;
K$ (covering a range of 2 $\sim$ 8 KeV).
The physics of the hot intercluster gas provides the means of
measuring directly the distance of clusters, without need of
intermediate steps. This method is based on the so called
{\em Sunyaev-Zeldovich effect} which is the distortion of the original
CMB spectrum, by the Compton scattering to higher energies of the CMB photons
from the hot electrons of the plasma (for recent reviews see
\cite{Rafae}, \cite{Birk}). This distortion decreases the
brightness of the CMB spectrum at the longer wavelength range while it
increases the photon energies in the shorter wavelength range. 
Lets assume a cluster of radius $R$ at a distance $D$, subtending an angular
separation $\theta$ on the sky. The change of the CMB brightness
temperature is proportional to
the line integral of electron number density through the cluster:
\begin{equation}
\frac{\Delta T}{T} \propto -\int T_\mathrm{e}(l) \; n_e \; \D
\end{equation}
then from isothermality we get $\Delta T/T \propto T_\mathrm{e} \; 
n_\mathrm{e} \; R$.
Furthermore the luminosity of the Bremsstrahlung radiation together
with the assumption of isothermality, gives:
\begin{equation}\label{eq:X1}
L_\mathrm{x} \propto \int n_\mathrm{e}^2 \; T_\mathrm{e}^{1/2}(r) \; 
4 \pi r^2 \; \D r = n_\mathrm{e}^2 \; T_\mathrm{e}^{1/2} \; R^3
\end{equation}
From (\ref{eq:X1}), the observed X-ray flux ($F_\mathrm{x} \propto 
L_\mathrm{x}/D^2$) and $R=\theta D$ we have:
\begin{equation}\label{eq:X2}
F_\mathrm{x} \propto n_\mathrm{e}^2 \; T_\mathrm{e}^{1/2} \; \theta^3 \; D 
\propto \left( \frac{\Delta T}{T} \right)^2 
\frac{\theta}{T_\mathrm{e}^{3/2} \; D}
\end{equation}
and solving for $D$ we get:
$$ D \propto \left( \frac{\Delta
T}{T} \right)^2 \frac{\theta}{T_\mathrm{e}^{3/2} F_\mathrm{x}}$$
Therefore measuring $F_\mathrm{x}$, $\theta$, 
$T_\mathrm{e}$ we obtain an absolute
determination of the cluster distance. The temperature, $T_\mathrm{e}$, can be
measured either from the shape of the X-ray continuum or from line
emission (especially of iron).
Note however, that we have
assumed sphericity, isothermality 
and a smooth distribution of $n_\mathrm{e}$. 
Most clusters are flattened (cf. \cite{BPM00}),
and show significant substructure apparent in
the optical and X-ray images. In such clusters the above
procedure may provide highly uncertain and biased distance estimates
(cf. \cite{Jet02}). A
recent study of $\sim 100$ clusters has shown that once cooling
flows are taken into account, isothermal profiles fit well 
$\sim 90\%$ of the clusters (\cite{Wh00}, but see also \cite{deG02}).

\subsection{The Value of $H_{\circ}$ and the Age of the Universe}
The immense effort that has been put towards the goal of
determining $H_{\circ}$, with an accuracy of a few \%, has recently
come to fruition with different methods providing consistent (within $<10\%$)
values. In Table 1, I present a list of some of the most recent determinations
of $H_{\circ}$. Note that there are 3 different measurements based on
SNIa, giving different values of $H_{\circ}$. Although many
of the SNIa they use are common, 
the difference is most probably attributable to
the different local calibrations that they employ. Thus, the
differences in their derived $H_{\circ}$'s should reflect the {\em systematic}
uncertainty introduced by the different local calibrations and it is
indeed comparable to the systematic uncertainty that the individual
studies have estimated.

The method in the last row of Table 1 is
based on {\em gravitational lensing} (cf. \cite{BN92}).
The basic principle behind this method is that there is a
difference in the light travel time along two distinct rays
from a source, which has been gravitationally lensed by some
intervening mass. The relative time delay ($\Delta t$), 
between two images of the source, can be measured if the
source is variable. Then it can be shown that the Hubble constant
is just:
\begin{equation}
H_{\circ} = {\cal C} \frac{\Delta\theta^2}{\Delta t}
\end{equation}
where $\Delta\theta$ is the image separation and ${\cal C}$ is a
constant that depends on the lens model. Although, this method has
well understood physical principles, still the details of the lensing
model provide quite large uncertainties in the derived $H_{\circ}$.

A crude $N$-weighted average of the different
$H_{\circ}$-determinations in Table 1 gives:
$$H_{\circ} = 68 \pm 6 \;\;\; \mbox{\rm km sec$^{-1}$ Mpc$^{-1}$}$$
where the uncertainty reflects that of the weighted mean (the
individual uncertainties have not been taken into
account).   However, there seems to be some clustering around two
preferred values ($H_{\circ} \simeq 60$ and $\simeq 72$ km s$^{-1}$
Mpc$^{-1}$) and thus the above averaging provides biased results. More
appropriate is to quote the median value and the 95\% confidence
limits:
\begin{equation}\label{eq:H_o}
H_{\circ} = 72^{+4}_{-13} \;\;\; \mbox{\rm km sec$^{-1}$ Mpc$^{-1}$} \;,
\end{equation}
The anisotropic errors reflect the non-Gaussian nature of the
distribution of the derived $H_{\circ}$-values. 
Note that the largest part of the individual
uncertainties, presented in Table 1, of all methods except the last
two, are {\em systematic} because they rely on local calibrators (like the
distance to the LMC), which then implies that a systematic offset of
the local zero-point will ``perpetuate'' to the secondary indicators although
internally they may be self-consistent. A further source of systematic
errors is the peculiar velocity model, used to correct the derived
distances, which can easily introduce $\sim 7\%$ shifts in the derived
$H_{\circ}$ values \cite{RR01} \cite{WilB}.
\begin{table}[t]
\caption{Some recent determinations of the Hubble 
constant, based on different methods.}
\centering
\tabcolsep 8pt
\begin{tabular}{lccccc} \hline
Method & N & $H_{\circ}$ & $\langle z \rangle$ & reference \\ \hline
Cepheid & 23 &75$\pm$10 & 0.006 & Freedman et al 2001 \\
2-ary methods & 77 & 72$\pm$8.0 &  $\lesssim 0.1$ & Freedman et al 2001 \\
IR SBF & 16 &76$\pm$6.2 & 0.020 & Jensen et al. 2000 \\
SN Ia & 36 &73$\pm$7.3 & $\lesssim 0.1$ & Gibson \& Brook 2001 \\
SN Ia & 35 & $59\pm6$ & $\lesssim 0.1$ & Parodi et al 2000 \\
SN Ia & 46 & $64\pm7$ & $\lesssim 0.1$ & Jua et al 1999 \\
CO-line T-F & 36 & 60$\pm$10 & $<0.11$ & Tutui et al 2001 \\
S-Z & 7 & 66$\pm$15 & $<0.1$ & Mason et al 2001\\
Grav. Lens & 5 & $68\pm13$ &  & Koopmans \& Fassnacht 2000 \\
\hline
\end{tabular}
\end{table}

With the value (\ref{eq:H_o}), we obtain a {\em Hubble} time, 
$t_\mathrm{H}$, equal to:
\begin{equation}\label{eq:Hof} 
t_\mathrm{H} = H_{\circ}^{-1} = 13.6^{+3}_{-0.6} \;\;\; {\rm Gyr's}
\end{equation}
the uncertainties reflecting the 95\% confidence interval. 

It is trivial to state that the present age of the Universe $t_{\circ}$,
should always be larger than the age of any extragalactic object.
A well known problem that has troubled cosmologists, is the fact that
the predicted age of the Universe, in the classical Einstein de-Sitter
Cosmological Model is smaller than the measured age of the oldest
globular clusters of the Galaxy. This can be clearly seen from
(\ref{eq:Hof}) and 
\begin{equation}\label{eq:nage}
t_{\circ}^\mathrm{EdS}=\frac{2}{3} 
t_\mathrm{H} \simeq 9^{+2}_{-0.7} \;\;\; {\rm Gyr's}
\end{equation}
and although the latest estimates of
the globular cluster ages have been drastically decreased to \cite{KC02}: 
$$t_\mathrm{gc} \simeq 12.5 \;\;\; \mbox{\rm with 95\% lower/upper limit is
11/16 Gyr's} \;.$$
One should then add the age of the formation of the globular
clusters and assuming a redshift of formation
$z\simeq 5$ then this age is $\sim 0.6 - 0.8$ Gyr's which brings the lower
95\% limit of $t_\mathrm{gc}$ to $\sim 11.6$ Gyr's (see however
\cite{BCla02} for possible formation at $z\gtrsim 10$).
It is evident that there is a discrepancy between $t_{\circ}$ 
and $t_\mathrm{gc}$. This discrepancy 
could however be bridged if one is willing to push in the right direction the
95\% limit of both $t_\mathrm{H}$ and $t_\mathrm{gc}$.

However, other lines of research point towards the age-problem.
For example, if at some large redshift we observe galaxies with old stellar
populations, for which we know the necessary time for evolution to
their locally ``present''state, then we can deduce again the age of the
Universe. In an EdS we have $R\propto t^{2/3}$ and thus we have:
\begin{equation}\label{eq:age2}
t_{\circ} = t_\mathrm{z} (1+z)^{3/2}
\end{equation}
Galaxies have been found at $z\simeq 3$ with spectra that
correspond to a stellar component as old as $\sim 1.5$ Gyr's, in their local
rest-frame. From (\ref{eq:age2}) we then have that $t_{\circ} \simeq 12$
Gyr's, in disagreement with the EdS age (\ref{eq:nage}).


This controversy could be solved in a number of ways, some of which are:
\begin{itemize}
\item invoking an open ($\Omega<1$) cosmological model, 
\item assuming that we live in a local underdensity of an EdS Universe,
\item invoking a flat model with $\Omega_\mathrm{\Lambda}>0$.
\end{itemize}

The first possibility is in contradiction with many observational data and
most importantly with the recent CMB experiments ({\sc boomerang, 
maxima} and {\sc dasi}), which show that $\Omega=1$
(see \cite{Bern00} \cite{Bern02} \cite{Lee} \cite{Stom} \cite{Pryk}).

The second possibility \cite{Tom} can solve the age-problem
by assuming that we live in a local underdense region that extends to
quite a large distance, which would then imply that the measured local
Hubble constant is an overestimate of the global one by a factor:
$$\frac{\delta H}{H} = \frac{\Omega_\mathrm{m}^{0.6}}{3 b}
\frac{\delta N}{N}$$
where the bias factor $b$, is the ratio of the fluctuations in
galaxies and mass. To reduce the Hubble constant to a
comfortable value to solve the age problem, say from
72 to 50 km sec$^{-1}$ Mpc$^{-1}$, one then needs $\delta N/N
\simeq -0.9/b$. Values of $b$ are highly uncertain and model dependent,
but most recent studies point to $b\sim 1$ (cf. \cite{Lah01})
which would then mean that we need to live in a local very underdense
region, something that is not supported by the linearity of the Hubble
relation out to $z\lesssim 0.03$ (cf. \cite{Giov99}) or out to
$z\lesssim 0.1$ (cf. \cite{Tam99}). This is not to say that we are not
possibly located in an underdense region, but rather that this cannot
be the sole cause of the {\em age}-problem \cite{Zehavi}.

Thus we are left with the last possibility of a Universe dominated 
by vacuum energy (a Universe with $\Omega_\mathrm{\Lambda}>0$ - 
see section 1.3). 
If we live in the accelerated phase (see Fig.1) we will measure:
$$H(t_{\circ}) > H(t) \;\;\;\;\;\; {\rm with} \;\; t_{\circ}>t$$
ie.,  a larger Hubble constant, and thus smaller Hubble time as we
progress in time, resolving the age-problem.
In fact we have strong indications (see next section), from the SNIa
results, which trace the Hubble relation at very large distances
(see section 3.2), and from the combined analysis of
CMB anisotropy and galaxy clustering measurements in the
2dF galaxy redshift survey \cite{Efst02},
for a flat Universe with $\Omega_\mathrm{\Lambda} \simeq 0.7$.
Then from (\ref{eq:lam}) we obtain the age of the Universe in such a
model:
$$ t^\mathrm{\Lambda}_{\circ} = 
1.446 \times t_{\circ}^\mathrm{EdS} = 13.1^{+2.8}_{-1.1} \; \mbox{Gyr's}$$ 
Indeed the resolution of the {\em age}-problem gives further support to the 
$\Omega_\mathrm{\Lambda}>0$ paradigm.

\newpage

\section{Determination of the Matter/Energy Density of the
Universe}
The existence of large amounts of Dark Matter in the universe,
manifesting itself through its gravitational effects, is a
well established fact, although the precise amount has been a matter
of a lively debate through the years. Attempts to identify the DM
as normal {\em baryonic} matter has failed, mostly due to the
nucleosynthesis constraints imposed by the successful hot Big-Bang
model and the
large temperature fluctuations of the CMB that 
it predicts in a flat Universe ($\frac{\delta T}{T} \sim \frac{1}{3}
\frac{\delta\rho}{\rho}$).
Possibly some of the DM could be neutral
hydrogen, in the form of Lyman-$\alpha$ clouds, but it is estimated
that it could contribute only $\Omega\lesssim 0.01$. Similarly, the
possible solid form of baryonic material 
(eg. dust grains, {\em Jupiters}, dwarfs with $M\lesssim 0.08
M_{\odot}$ or neutron stars) would contribute little to $\Omega$.

Two recent determinations of the deuterium abundance, which combined
with the BBN (Big-Bang Nucleosynthesis) predictions, give 
the total baryonic mass in the
universe, have provided slightly discrepant results (see reviews
\cite{OSW}, \cite{Stei}), covering the range:
\begin{equation}\label{eq:BBN}
0.005 \lesssim \Omega_\mathrm{B} h^{2} \lesssim 0.024
\end{equation}
and therefore for $h=0.72$ we have: $ 0.01 \lesssim \Omega_\mathrm{B} 
\lesssim 0.046$. However, the recent analysis of the results from the
{\sc boomerang} CMB experiment have provided a value mostly compatible
with the lower deuterium abundance and thus higher $\Omega_\mathrm{B}
h^{2}$ value (see further below).

In this section we present a variety of methods used
to estimate either the total mass/energy density of the
Universe or its mass density, $\Omega_\mathrm{m}$.
\begin{figure}[t]
\includegraphics[width=12cm,height=9cm]{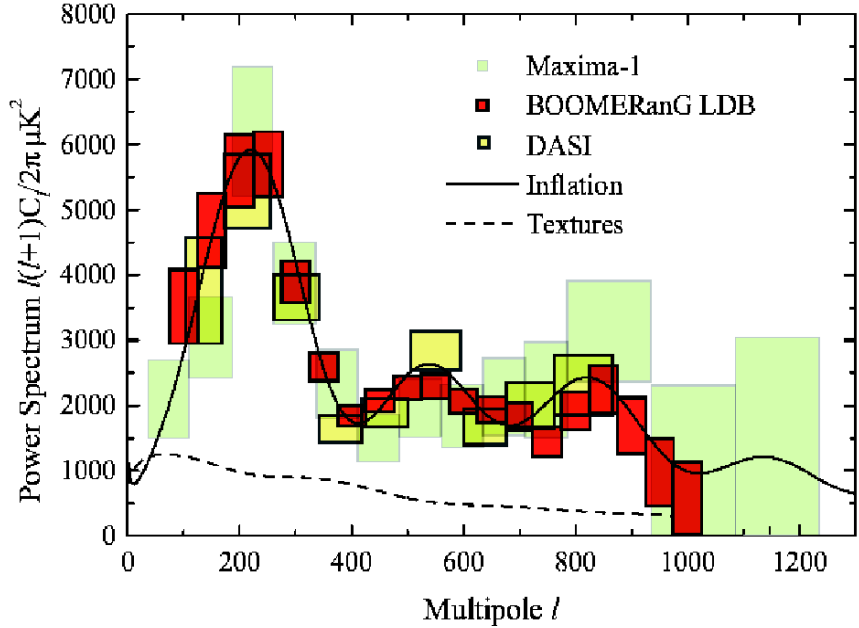}
\caption{CMB spectrum from the {\sc bommerang, maxima} and {\sc dasi} 
experiments
with the error boxes of the measurements.
The predictions of the popular inflationary model and
one non-Gaussian (global texture) model (from \cite{Melc02} with permission).}
\end{figure}

\subsection{The CMB fluctuation spectrum}
The most straight forward approach to estimate the total matter and energy 
density of the Universe (ie., the total $\Omega$) is by means of the
measurement of the fluctuation spectrum of the CMB.
Before recombination at $z\lesssim 1100$, the
baryons and photons are tightly coupled, oscillating acoustically due
to gravity (on sub-horizon scales). Only after recombination do the
acoustic oscillations stop and the 
density fluctuations grow. 
The fluctuations emerging from the last scattering surface are a
series of peaks and troughs \cite{SZ70} and as the
different wave-lengths are projected to different angular scales on
the last scattering surface and depending on the underlying 
cosmological model, they produce a characteristic structure of
peaks on the CMB power spectrum (for a recent review see \cite{Melc02}
and references therein). This method is in effect based in
measuring the angular extent of a physical scale on the last
scattering surface. The curvature of space enters through the
angular distance to the last scattering surface. Therefore, the same physical
scale will be projected to a smaller angular scale on the CMB sky in a
positively curved background, while it will be projected to a
larger angular scale in a flat or to an even larger scale
in a negatively curved background space.
\begin{figure}[t]
\includegraphics[width=9cm,height=9cm,angle=-90]{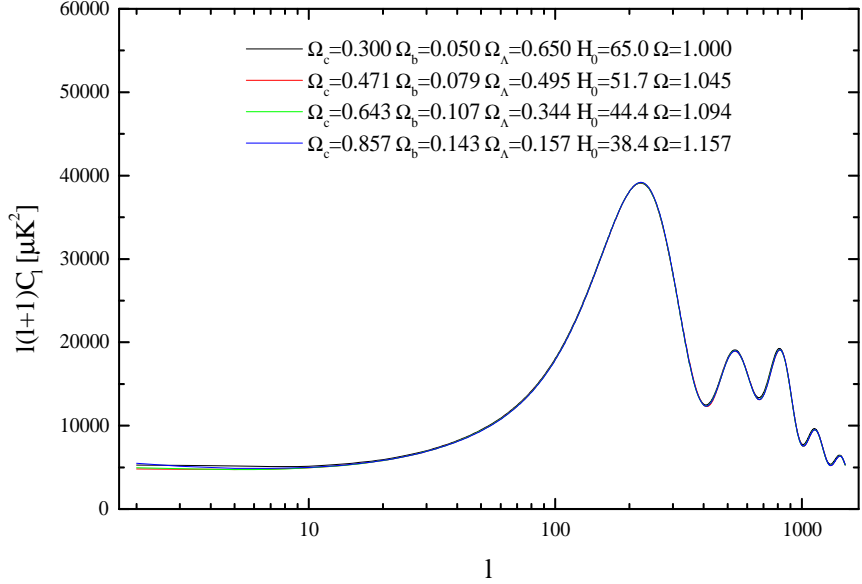}
\caption{Different combinations of the cosmological parameters can
result in the same CMB power-spectrum - degeneracy problem (form
\cite{MG01} with permission)}.
\end{figure}

To define the CMB power spectrum one starts by expanding
the temperature fluctuations of the CMB sky in spherical harmonics:
\begin{equation}
\frac{\delta T}{T}(\theta,\phi) = \sum_{\ell=1}^\infty\sum_{m=-\ell}^{m=\ell}
	 a_{\ell m} Y_{\ell m}(\theta,\phi)~,
\end{equation}
then if the fluctuations are Gaussian, the 2-point 
correlation function contains all the statistical information, and 
can be defined as:
\begin{equation}
\left\langle \left(\frac{\delta T}{T}\right)^{2} \right\rangle =
 {1\over 4\pi} \sum_\ell (2\ell+1)\; C_\ell \; W_\ell \; P_\ell(\theta,\phi)~.
\end{equation}
where $W_\ell$ is the window function representing the beam
characteristics of the experimental apparatus used to observe the CMB
sky, while the average is over all positions on the sky. One then invokes
the ergotic theorem, ie, that the above average is equivalent to being
over different realizations of our Universe.
Then assuming random phases one can define the CMB power
spectrum $C_\ell$ as the ensemble average of the coefficients
$a_{\ell m}$:
$$C_\ell = \langle|a_{\ell m}|^2\rangle$$

The different cosmological parameters will reflect onto a different
structure of peaks in the structure of the CMB power spectrum. 
The position of the first 
peak is determined by the global mass/energy 
density of the Universe and the dependence of $\ell_\mathrm{peak}$ on 
$\Omega$ can be approximated by:
\begin{equation}
\ell_\mathrm{peak}\sim{220 \over \sqrt{\Omega}}
\end{equation}
Note however, that this approximation is not correct in $\Lambda$--dominated 
universes and small corrections should be applied
(cf. \cite{MG01}). 
Many recent experiments like the {\sc bommerang, maxima} and
{\sc dasi} (cf. \cite{Bern00}, \cite{Bern02}, \cite{Lee} \cite{Stom}
\cite{Pryk}) find:
$$\Omega_\mathrm{tot} \simeq 1.02 \; (\pm 0.07) $$

Many other cosmological parameters (for example $\Omega_\mathrm{m}$, 
$\Omega_\mathrm{\Lambda}$, $H_{\circ}$, baryon content of the universe, the
spectral index $n$ of the inflationary perturbation spectrum, etc)
affect the structure of the peaks, beyond the first one
(cf. \cite{Hu97}).  Determining
the CMB spectrum up to a few thousand $\ell$'s can 
put strong constraints on these parameters. Current experiments
trace the CMB spectrum up to $\ell \sim 1000$ and indeed
they have detected two more significant 
peaks at roughly $\ell \sim 540$ and 840 \cite{Bern02} (see Fig.6). 

Note however, that 
different combinations of the cosmological parameters can conspire to
produce exactly the same CMB spectrum; this is the so called
degeneracy problem (see Fig.7) and therefore in order to provide
limits to these cosmological parameters one needs to assume
priors and/or constrain different combinations of these parameters. 
However, the more accurate the derived CMB spectrum the weaker
the necessary priors\footnote{With the new CMB experiments - {\sc MAP} 
and {\sc PLANCK} 
- the CMB power spectrum will be determined to an unprecedent detail,
providing extremely accurate values for more than 10 cosmological
parameters \cite{Smoot}.}.

The latest data and CMB spectrum analysis
provides very stringent constraints to the baryon content of the
Universe: $\Omega_\mathrm{B} h^{2} \simeq 0.022^{+0.004}_{-0.003}$,
consistent with the primordial nucleosynthesis 
constraints (see \ref{eq:BBN}), and to the
spectral index of the power spectrum of primordial
perturbations: $n \simeq 0.96 \pm 0.1$ \cite{Bern02}.
Furthermore, combined analyses with other cosmological data, can be
used to break the above mentioned degenerecies (see below).

\subsection{The Hubble diagram with SNIa}
As we have already discussed in section 2.3, the Hubble diagram of
supernovae SNIa can be used not only to determine the Hubble constant (at
relatively low redshifts) but also to trace the curvature of the
Hubble relation at high redshifts (see \cite{Ries00} and
references therein).
\begin{figure}[t]
\centering{
\includegraphics[width=9.cm,height=8.cm]{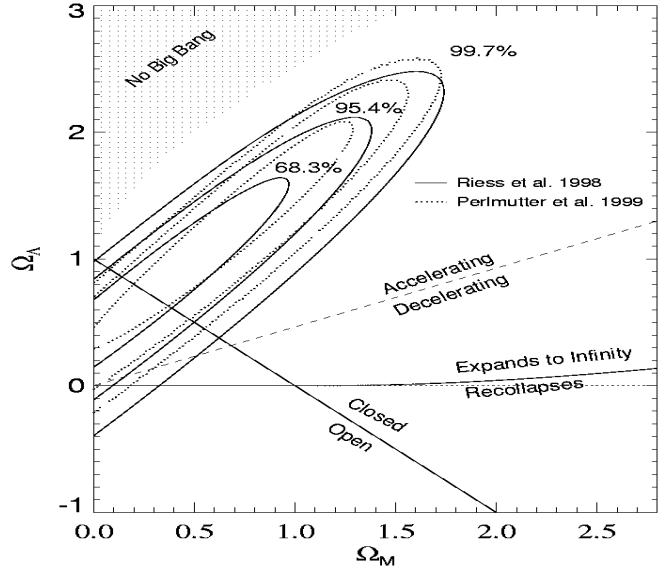}
}
\caption{Confidence intervals for $(\Omega_\mathrm{m} - 
\Omega_\mathrm{\Lambda})$ from
the SCP and HZT results (from \cite{Ries00} with permission).}
\end{figure}

The two groups working laboriously on this subject
(SCP and HZT) have found consistent results, by which the distant
SNIa's are dimmer on average by 0.2 mag than what expected 
in a flat EdS model, which translates in them being $\sim 10\%$ further
away than expected (\cite{Perl99}, \cite{Ries98}).
This implies that we live in an accelerating phase
of the expansion of the Universe, a fact that supports
a non-zero cosmological constant. The confidence intervals that
their results put in the $\Omega_m - \Omega_\mathrm{\Lambda}$ plane are shown
in Fig.8.
\begin{figure}[t]
\centering{
\mbox{\epsfxsize=10cm \epsffile{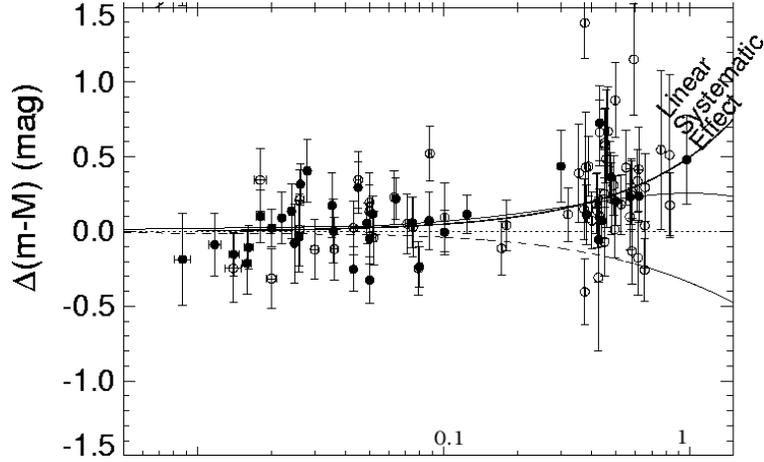}}}
\caption{Distance modulus residuals after subtracting an open
$\Omega_m=0.3$ Hubble relation (straight dashed line). The flat 
$\Omega_{\Lambda}=0.7$ model is the thin curved line while the
systematic effect is the thick label line (from \cite{Ries00} with 
permission).}
\end{figure}
These results can be quantified by the following expression \cite{Perl99}:
$$0.8\Omega_\mathrm{m} - 0.6\Omega_\mathrm{\Lambda}=0.2 \; (\pm 0.1)$$
Together with the CMB fluctuation spectrum results we obtain:
$$\Omega_\mathrm{m} \simeq 0.29 \;\;\;\; \Omega_\mathrm{\Lambda}\simeq 
0.71\;.$$
However, since our understanding of the physics of SNIa'a is not
complete (cf. \cite{Liv}, \cite{Leib}) there could be some
systematic effect, correlated with distance (eg. evolution), which could
explain the dimming of the distant SNIa's and thus alleviate the
$\Omega_\mathrm{\Lambda} > 0$ interpretation. In Fig.9 we show the
distance modulus residuals after subtracting an open
$\Omega_\mathrm{m}=0.3$ Hubble relation. 
The systematic distant-dependent effect 
mimics the accelerated expansion Hubble relation out to $z\sim 0.8 -
1$. Beyond $z\sim 1$ the two relations depart due to the fact that the
accelerated phase has to first pass from a decelerating one (see
discussion in section 1.3) and this could provide a
strong test for the possible distant dependent systematics.
In fact, the recent discovery of the furthest known supernova (SN 1997ff) at a
redshift of $z\sim 1.7$ \cite{Ries01},
has provided evidence of the decelerating
phase of the presently accelerating Universe (however, more very
high-$z$ supernovae are necessary to confirm this extraordinary result).

\subsection{Clustering of Galaxies, Clusters \& QSO's}
If we assume a continuous density field of extragalactic objects, 
$\rho(\vec{r})$, with mean $\overline{\rho}$, we define the
fluctuations of the field at position $\vec{x}$ as:
\begin{equation}\label{eq:fluct}
\delta(\vec{x}) = \frac{\rho(\vec{x}) - \overline{\rho}}{\overline{\rho}} 
\;,
\end{equation}
Obviously we have that $\langle \delta(\vec{x}) \rangle = 0$. 
The correlation function is defined as:
\begin{equation}\label{eq:cont}
\xi(r) = \langle \delta(\vec{x}) \delta(\vec{x} + \vec{r}) \rangle =
\frac{\langle \rho(\vec{x} + \vec{r}) \rho(\vec{x}) 
\rangle - \overline{\rho}^{2}}{\overline{\rho}^{2}} \;.
\end{equation}
and quantifies the extend to which the density 
fluctuations at a given point are correlated to those at a distance $r$.
The value of 2-point correlation function at zero-lag
is therefore the variance of the random process:
$$\xi(0) = \langle \delta^{2}(\vec{x}) \rangle$$
which measures the excursions of the density about its mean value. However, 
this is not a well defined quantity because usually the fluctuation field
is smoothed to some resolution, say $R$.
We then evaluate the variance of this field 
as $r\longrightarrow 0$ (see \ref{eq:var} below).

In many problems it is convenient to work in wave-number space.
The Fourier transform of $\delta(\vec{x})$ is:
\begin{equation}
\delta_\mathrm{\vec{k}} = \int \delta(\vec{x}) e^{i \vec{k}\cdot
\vec{x}} \D^{3}\vec{x} 
\end{equation}
and it is convenient to separate $\delta_\mathrm{\vec{k}}$ in 
modulus and argument:
$$\delta_\mathrm{\vec{k}} = |\delta_\mathrm{\vec{k}}| 
e^{i\epsilon_\mathrm{\vec{k}}} \;, $$
where $\epsilon_\mathrm{\vec{k}}$ are the phases (usually assumed to be 
randomly 
distributed in $[0,\pi)$, although the non-linear evolution of structure
introduces phase correlations, cf. \cite{phase}).
The variance of the amplitudes is the power spectrum:
$$P(\vec{k}) \equiv \langle|\delta_\mathrm{\vec{k}}|^{2}\rangle$$
which is the Fourier transform of the correlation
function ({\em Wiener-Khinchin} theorem):
\begin{equation}
P(\vec{k}) = \int \xi(\vec{r}) e^{i \vec{k} \cdot \vec{r}} \;\D^{3}\vec{r}
\end{equation}
and with inverse transform:
$\xi(\vec{r}) = (2\pi)^{-3} \int P(\vec{k}) e^{-i 
\vec{k}\cdot \vec{r}} \D^{3}\vec{k}$.
At the origin $\vec{r} = 0$ we obtain
\begin{equation}\label{eq:var}
\xi_\mathrm{R}(0) = \langle \delta_\mathrm{R}^{2}(\vec{x}) \rangle = 
\frac{1}{(2\pi)^{3}} \int_\mathrm{\vec{k}} P(\vec{k}) \; W^{2}(kR) 
\D^{3}\vec{k}
\end{equation}
where $W$ is the window function function that reflects the filtering
of the field. The power spectrum is the contribution of modes of
wavenumber $\vec{k}$ to the total variance, per unit volume of wavenumber 
space. If the fluctuation field is a {\em Gaussian Random Field}, then
the power-spectrum contains all the statistical information of the
fluctuations. 
A similar formulation is applicable also in the two dimensional case (where
the density field is on the surface of a sphere - the sky). Only that instead
of the Fourier transform we use the Spherical Harmonic transform.

Although I will not enter in the details of how to estimate the power
spectrum of some distribution of extragalactic objects, I will only
note that a good estimation of the window function (containing the
survey boundaries, obscuration, radial selection function and 
instrumental biases for example) 
is necessary in order to get a reliable power-spectrum determination.

A further ingredient that complicates considerably matters is
the so called biasing of the galaxies, or in general of any
mass-tracer population, with respect to
mass \cite{Kaiser}. The usual relation, assumed between the mass-tracer 
fluctuations ($\delta_\mathrm{tr}$) and the underline 
mass fluctuation field, is encapsulated in the bias factor $b$:
\begin{equation}\label{eq:bias}
\delta_\mathrm{tr}= b \; \delta_\mathrm{m}
\end{equation}
and therefore we have that the galaxy (tracer) power spectrum is 
$$ P_\mathrm{tr}(k) = b^{2} P_\mathrm{m}(k)$$
It has been shown that the linear biasing model (\ref{eq:bias}) is a
good approximation, at least on scales were non-linear gravitational
effects are weak (see \cite{Muns} and references therein).

Within the inflationary paradigm the initial fluctuations, in the
early universe, that gave rise to the observed large-scale structure
today, are {\em adiabatic} and {\em Gaussian} and therefore  
one can characterize these fluctuations completely using the above
tools. The power spectrum of such initial fluctuations is:
$$P_\mathrm{in}(k) = A k^{n}$$
usually with $n=1$ ({\em Harrison-Zeldovich} spectrum), and $A$ its
amplitude. 
The different fluctuation damping mechanisms, operating during 
the radiation dominated area, modify $P_\mathrm{in}(k)$. 
These effects can be encapsulated
in the {\em transfer function}, $T(k)$ and today's linear fluctuation
spectrum has the form \cite{Bar86}:
$$P(k) = T^2(k) P_\mathrm{in}(k)$$
In the linear regime (while fluctuations $\ll 1$) the power spectrum
shape is preserved, because each Fourier mode evolves
independently. 

I will now concentrate on a few methods, based on the clustering of
galaxies and QSO's, and I will present only
very recent results. The subject is extremely rich, many have
laboriously worked towards attaining the goal of pining down the
different cosmological parameters and I hope that they will forgive me for
not being able to mention the vast literature on the subject. 

\subsubsection{The Shape of $P(k)$:}
The popular Cold Dark Matter (CDM) model has a $T(k)$ parametrised by
the so-called {\em shape-parameter}, $\Gamma$, which characterizes the
shape of the $P(k)$ and has the form \cite{Sugi}:
\begin{equation}
\Gamma=\Omega_\mathrm{m} h \;\exp{\left[-\Omega_\mathrm{B} 
(1+\sqrt{2 \;h}/\Omega_\mathrm{m}) \right]}
\end{equation}
We see that measuring the the power-spectrum of extragalactic populations
and estimating $\Gamma$ we can put constraints on the combination of
the cosmological parameters: $\Omega_\mathrm{m}$ and $h$
($\Omega_\mathrm{B}$ affects weakly $\Gamma$).

The recent 2dF survey \cite{2df} has measured already measured more than 160000
galaxy and 10000 QSO redshifts, 
which constitute it the largest spectroscopic catalogue of
extragalactic objects. 
\begin{figure}[t]
\centering{
\includegraphics[width=10.cm,height=8.cm]{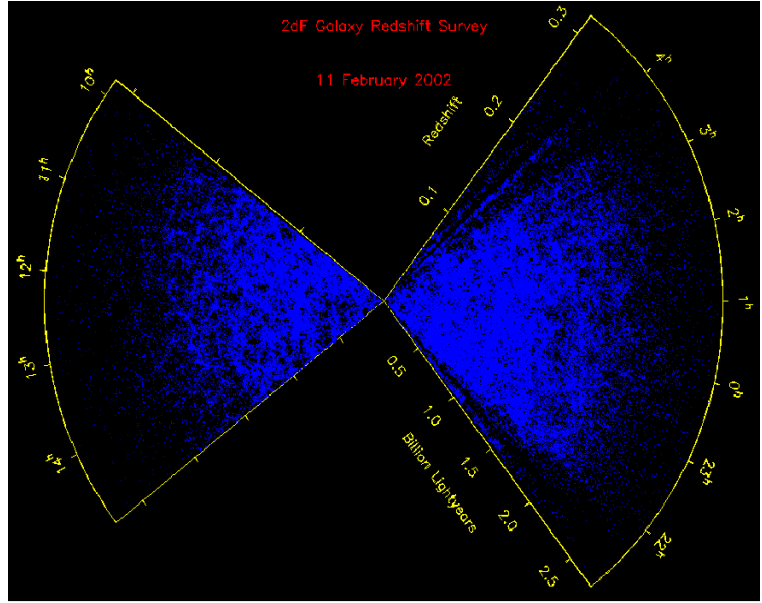}
}
\caption{Pie-diagram of the latest 2dF release, containing 160000
galaxies (with permission from the {\sc 2dfgrs team}).}
\end{figure}
A number of recent papers have estimated the galaxy and QSO 
power-spectra providing important constraints on the $\Gamma$ parameter.
From the $P(k)$ of galaxies \cite{Perci} it was found that there is a
degeneracy between the $\Omega_\mathrm{m} h$ and 
$\Omega_\mathrm{B}/\Omega_\mathrm{m}$, which if broken (using the CMB
or BBN results) provide the following constraints:
$$\Omega_\mathrm{m} \;h \simeq 0.2 (\pm 0.03) \;\;\;\;
\Omega_\mathrm{B}\simeq 0.15 (\pm 0.07) \Omega_\mathrm{m}$$
which for $h=0.72$ means that $\Omega_\mathrm{m}=0.28$.
The analysis of the QSO $P(k)$ \cite{Hoyle} showed a
somewhat smaller value but still consistent within the 
statistical uncertainties:
$$\Gamma \simeq 0.1 (\pm 0.1)$$
Similar results come from the {\sc ssds} photometric survey
\cite{ssds} which
contains 1.5$\times 10^{6}$ galaxies with redshifts up to $\sim 0.4$. 
The angular power-spectrum analysis, after inverting to 3D 
using Limber's integral
equation, give \cite{Dodel} \cite{Szal}:
$$\Gamma \simeq 0.14^{+0.11}_{-0.06} \;\; \mbox{95\% C.L.}$$ 
Obviously, all these results support a low $\Omega_\mathrm{m}$ universe.

\subsubsection{Redshift Space Distortions:}
As we have already discussed in section 2, the measured expansion
velocity of an extragalactic object contains also the contribution of
the local gravitational field. The peculiar
velocities will distort the apparent 3D distribution of the
extragalactic objects, a fact which will manifest itself in the
2-point correlation function, when plotted as a function of the transverse and
radial pair separation. The redshift space correlation function is
related to the real space, under a few assumptions; see \cite{Kais87},
according to:
\begin{equation}\label{eq:xired}
\xi_\mathrm{s}(r) = \xi_\mathrm{r}(r) \left(1+\frac{2}{3} 
\beta+\frac{1}{5} \beta^2\right)
\end{equation}
where $\beta = \Omega_\mathrm{m}^{0.6}/b$.
By estimating the angular power-spectrum, $C_\ell$, or its Fourier
transform $w(\theta)$, and 
then inverting it to 3D via Limber's equation, one has  
an unaffected, by redshift space distortions, measure of
these parameters. Then using (\ref{eq:xired}) and the measured 
$\xi_\mathrm{s}(r)$ one can place constraints on $\beta$. Such an
analysis of the 2dF galaxy survey gave \cite{Peac_2}:
$$\Omega_\mathrm{m} \simeq 0.25 (\pm 0.06) \; b_\mathrm{op}^{-1.66}$$
A subsequent analysis, using a different method, provided very similar
results \cite{Tegm}:
$$\Omega_\mathrm{m} \simeq 0.3 (\pm 0.15) \; b_\mathrm{op}^{-1.66}$$
The corresponding QSO survey \cite{Outram} did not provide very stringent
constraints (ie., the EdS model was rejected only at a 1.4$\sigma$
level), however their best fit gives:
$$\Omega_\mathrm{m} \simeq 0.21 (\pm 0.15) \; b_\mathrm{QSO}^{-1.66}$$

\subsubsection{Joint Likelihoods:}
Alot of recent interest was generated by the understanding that
joining the analyses of different data sets, one may break the
degeneracies between the different cosmological parameters (cf.
\cite{Webster} \cite{Efst99} \cite{Brid01} \cite{Wang}).
Especially joining the CMB, SNIa and large-scale clustering results may
lead to strong constraints on more than 8 cosmological
parameters (see however \cite{Lah_s} for many subtleties involved).

The joint analysis \cite{Efst02} of the  2dF 
galaxy $P(k)$ and the CMB data
have provided another strong
indication for a positive cosmological constant, 
independent of the SNIa results, with a 2$\sigma$ range:
$$0.65 < \Omega_\mathrm{\Lambda} < 0.85$$ 
Furthermore, some of the other constraints are:
$0.17 < \Omega_\mathrm{m} <0.31 \;\;\; \mbox{for} \; h=0.72$, 
$0.6 < h < 0.86$.

\subsection{$M/L$ observations}
Each astronomical
object is characterized by a quantity called Mass-to-Light ratio, $M/L$.
A convenient scaling of $M/L$ is done by using the 
value of the solar neighbourhood, $M_{\odot}/L_{\odot}$. Then,
observing $M/L > M_{\odot}/L_{\odot}$ would imply
the existence of 
DM of unknown composition and origin. Note that in most cases
the evidence for $M/L > M_{\odot}/L_{\odot}$ comes from the gravitational 
effects of the DM.
Different classes of extragalactic objects (galaxies, clusters, etc)
are characterized by different $M/L$, indicating that possibly a
different fraction of the total mass of each type of object is {\em
Dark}.

Estimating the universal luminosity density (see further below) 
and 
using the derived $M/L$ values of each class of extragalactic object
we can estimate its contribution to the total 
$\Omega_\mathrm{m}$. Furthermore, if
the estimated $M/L$ is representative of the global universal value,
then we can derive the overall value of $\Omega_\mathrm{m}$.

I will present below the basic ideas behind the determination of $M/L$
for the different extragalactic populations.
 
\subsubsection{Spiral Galaxies:}
The Rotation curves of spirals (see Fig.10) are obtained by
measuring Doppler-shifts of emission lines in HII regions, at radio 
wavelengths using the 21-cm emission line of neutral Hydrogen or using
the CO-line and in
the latter cases the rotation curve is measured at several times the
optical radius of a galaxy (for a recent review see \cite{SR01}).

One would expect the rotation curve to fall roughly as the surface brightness
and beyond a few length-scales to fall as $v_\mathrm{rot} 
\propto r^{-1/2}$ (because 
most of the mass is rather centrally located) which is not observed. 
The rotation curves are found to be flat as far as 
they can be observed. From simple Newtonian Physics we have that:
$$ v_\mathrm{rot} \approx \left(\frac{G M}{r} \right)^{1/2}$$
and since $v_\mathrm{rot} \propto$ constant, we have that $M(r)
\propto r$, ie., mass increases linearly with distance beyond its 
optical radius, an indication for the presence of {\em dark matter}.
The average value of $M/L$ found for spiral galaxies out to $\sim 20$
kpc, is:
\begin{equation}
M/L \sim 10 \pm 2 M_{\odot}/L_{\odot}
\end{equation}
\begin{figure}[t]
\centering{
\mbox{\epsfxsize=12cm \epsffile{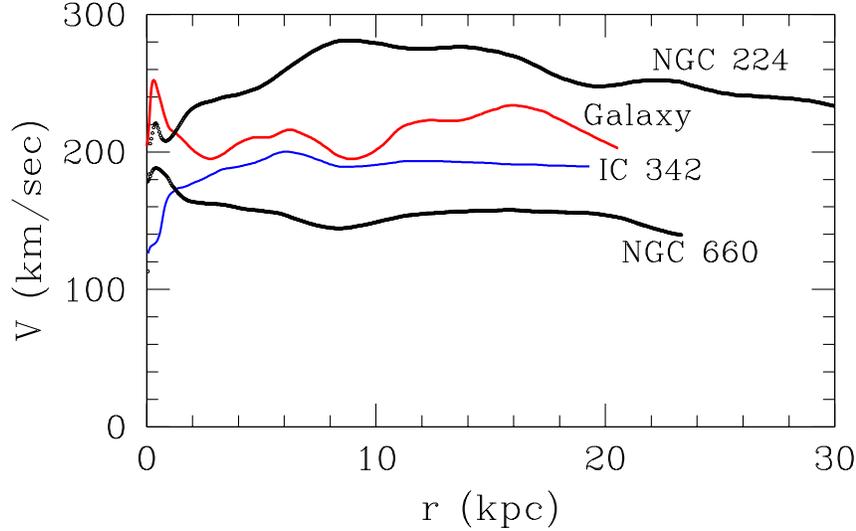}}}
\caption{Rotation curves of 4 spiral galaxies, one of which is the
Galaxy (data taken from \cite{Sof97}).}
\end{figure}
Note that the $M/L$ value is an increasing function of outer radius,
implying the existence of an extended dark matter halo (cf. \cite{BLD95}).

\subsubsection{Elliptical Galaxies:}
In principle one can invoke a similar method with that of the spirals,
if instead of the rotational velocity, the stellar velocity dispersion
is used. The amplitude of this velocity measure is dictated by the
gravitational potential of the elliptical and thus the
virial theorem can be used to determine its total gravitating mass.
However, stellar orbits are highly uncertain and 
the velocity dispersion ellipsoid may not be isotropic (see discussion
in \cite{Binney}).

Luckily, elliptical galaxies contain very hot gas 
($T\sim 5 \times 10^{6}$ $K$) which emits 
X-rays. This gas constitutes about $10\%$ of the observed mass in stars.
For a spherically symmetric galaxy in hydrostatic equilibrium ($\vec{v}=0$)
we have from {\em Euler's} equation that:
$$\frac{\partial \vec{v}}{\partial t} + \left( \vec{v} \cdot \nabla \right) 
\vec{v} = -\frac{1}{\rho} \nabla p - \nabla \Phi \Longrightarrow$$
$$ \frac{\D p}{\D r} = -\frac{G M(\le r) \rho}{r^{2}} \;,$$
Using the ideal gas law ($p =\rho k_\mathrm{B} T/ m$) 
with $m$ the molecular mass and $k_\mathrm{B}$ the Boltzmann constant, we obtain that:
\begin{equation}\label{eq:Tgr}
M(\le r) = \frac{k_\mathrm{B} T r}{\mu m_\mathrm{p} G} 
\left( -\frac{\D \ln \rho}{\D \ln r} - 
\frac{\D \ln T}{\D \ln r} \right)
\end{equation}
where $T$ is the gas temperature, $m_\mathrm{p}$ 
is the proton mass and $\mu$ is the 
mean molecular weight. Therefore if we measure the temperature and density 
profiles we can find the total mass distribution $M(<r)$. 
Finally the average mass-to-light ratio from this class of objects,
out to $\sim 20$ kpc, is:
\begin{equation}
M/L \sim 25 \pm 5 M_{\odot}/L_{\odot} \;.
\end{equation}

\subsubsection{Groups of Galaxies:}
Groups of galaxies containing a few galaxies (3 - 10) are usually considered 
bound (due to the high density of galaxies). Then, according to the Virial 
theorem ($2T + U=0$), a group with $N$ galaxies having velocity and position
vectors (relative to the centre of mass of the group) $\vec{v}_\mathrm{i}$ and 
$\vec{r}_\mathrm{i}$ respectively, has:
$$ \sum_\mathrm{i=1}^\mathrm{N} m_\mathrm{i} v_\mathrm{i}^{2} = 
\sum_\mathrm{i=1}^\mathrm{N} \sum_\mathrm{j<i} \frac{G m_\mathrm{i} 
m_\mathrm{j}}{|\vec{r}_\mathrm{i} - \vec{r}_\mathrm{j}|} \;.$$
However what we measure is not $v_\mathrm{i}$ 
but only the line-of-sight component 
of the the velocity $u_\mathrm{i}$. Assuming isotropic orbits then
$ v_\mathrm{i}^{2} \simeq 3 \langle u^{2}_\mathrm{i} \rangle$.
Assuming that the mass-to-light ratio of each member galaxy is the
same, $ m_\mathrm{i}/L_\mathrm{i} = M/L$, we obtain:
\begin{equation}
\frac{M}{L} = \frac{3\pi}{2G} \frac{\sum L_\mathrm{i}  u^{2}_\mathrm{i}}
{\sum \sum_\mathrm{j<i} L_\mathrm{i} L_\mathrm{j}  
|\vec{R}_\mathrm{i} - \vec{R}_\mathrm{j}|^{-1} }
\end{equation}
where $|\vec{R}_\mathrm{i} - \vec{R}_\mathrm{j}|$ 
is the projection of $|\vec{r}_\mathrm{i} 
- \vec{r}_\mathrm{j}|$ on the plane of the sky.

However this measure should be used only if $N$ is large, or otherwise
statistically, ie, averaged over
a large number of groups to find the $\langle M/L \rangle$ of the ensemble.
Note that this estimator is very sensitive to the 
group-finding algorithm, selection procedure, selection function
corrections (ie., corrections to take into account 
the fact that the galaxy density 
artificially decreases with depth in magnitude/flux limited samples). 
Furthermore, projection effects (ie; 
inclusion in the group of a foreground or background galaxy) is also a 
source of error. For example the inclusion of a non-member would result in an 
artificially higher $M/L$ while conversely the exclusion of a member (due to 
a high $u_\mathrm{i}$) would underestimate the $M/L$.

The mass-to-light ratio for groups of galaxies (cf. \cite{Ramel89})
is:
\begin{equation}
M/L \sim 180 \pm 50 M_{\odot}/L_{\odot} \;.
\end{equation}

\subsubsection{Clusters of Galaxies:}
The most common approach to measure the cluster $M/L$ is based on the
assumption that clusters are in virial equilibrium, for which
we do have strong indications.
\begin{figure}[t]
\centering{
\mbox{\epsfxsize=12cm \epsffile{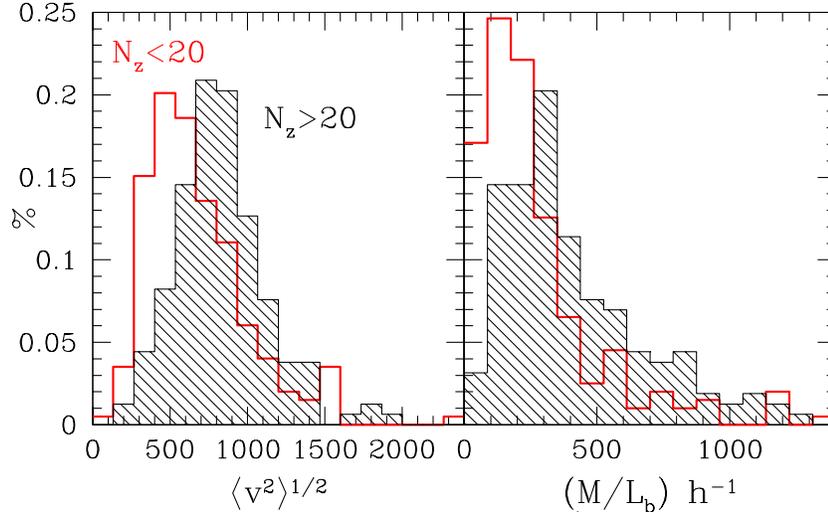}}}
\caption{Left panel: Frequency distribution of {\sc abell/aco} 
cluster velocity
dispersions based on $N_\mathrm{z}<20$ (empty histogram) and on 
$N_\mathrm{z}>20$ (shaded histogram).
Right panel: The corresponding $M/L$ distribution.}
\end{figure}
The first step in determining $M/L$ is to measure the cluster velocity
dispersion accurately. Using the recent compilation 
of 395 $R\ge 0$ {\sc abell/aco} 
cluster velocity dispersions \cite{Str99} I find:
$$ \sigma \equiv \langle (v-\bar{v})^{2} \rangle^{1/2} \approx 750 \pm 322 
\; \mbox{km/sec}$$
However, this value is based even on clusters for which the velocity
dispersion was determined from a very small number of galaxy
redshifts, a fact which increases the possibility of assigning an
erroneous $\sigma$-value. 
This can be seen in the left panel of 
Fig.11 were we plot the normalized $\sigma$ 
frequency distribution for clusters with $N_\mathrm{z}<20$ and 
$N_\mathrm{z}>20$. 
The skewed distribution to large $\sigma$-values for $N_\mathrm{z}<20$ 
is evident, while the mean $\sigma$-value is smaller than for the 
$N_\mathrm{z}>20$ case.
As a compromise between
having enough redshifts per cluster, in order to get a reliable $\sigma$-value,
and a large cluster sample, we choose
those with $N_\mathrm{z}>20$.  For these 195 clusters I find that:
$$ \sigma \approx 822 \pm 294 \; \mbox{km/sec} \;.$$
A consistent value of $\sigma \simeq 940 \pm 208$ km/sec has been found 
from the 16 high-redshift clusters of the CNOC
project \cite{Carl97} and a variety of methods used to define
cluster membership and account for interlopers \cite{Borg99}.

Now from the Virial theorem, $M=2 \sigma^2 r_a / G$, we have:
\begin{equation}
M_\mathrm{c}(\le r_\mathrm{a}) \approx 7 \times 10^{14} \; 
\left(\frac{\sigma}{1000}\right)^{2} \; h^{-1}  \; M_{\odot} \;.
\end{equation}
where $r_\mathrm{a}=1.5 \; h^{-1}$ Mpc. Using a number-galaxy weighted
luminosity estimation, and assuming
an average cluster value in the optical of $L \simeq 10^{12} \; h^{-2} \; 
L_{\odot}$ (cf. \cite{Bac99}), we obtain the distribution of
$M/L$ values (see right panel of Fig.11). Since the
distribution is non-Gaussian we quote below 
the median and 68\% confidence levels:
\begin{equation}\label{eq:ML_c}
(M/L)_\mathrm{c} \approx 320^{+170}_{-85} \; h \; M_{\odot}/L_{\odot}
\end{equation}

A more accurate method that can be used to estimate cluster mass
is based on the measurements of the X-ray emission from the ICM gas, 
similar to the method  used for individual elliptical galaxies 
(\ref{eq:Tgr}).
Although not strictly correct, usually an isothermal cluster profile is
used ($\D \ln T/ \D \ln r=0$), which greatly simplifies
calculations. However, recent experiments have shown that indeed the
cluster Temperature does not vary significantly with radius 
($\D \ln T/ \D \ln r\simeq -0.7 \sim -0.8$). Estimates of $M/L$
derived with this method are in general agreement with (\ref{eq:ML_c}).

\subsubsection{Global luminosity density and $\Omega_\mathrm{m}$:}
We estimate the value of $\langle L \rangle$ using the galaxy
luminosity function which is defined such that $\Phi(L) dL$ is the 
number density of galaxies having total luminosity in the interval
$(L, L+dL)$. A good fit to the observed luminosity
function of the field population of galaxies 
is provided by the {\em Schechter} function \cite{Sch76}:
\begin{equation}\label{eq:lum}
\Phi(L) = \frac{\phi_{*}}{L_{*}} \left(\frac{L}{L_{*}} \right)^{\alpha}
\exp{[-L/L_{*}]}
\end{equation}
where $\alpha = -1.29 \pm 0.11$, $ \phi_{*} = 1.3 \pm 0.3 
\times 10^{-2} h^{3} \mbox{ Mpc$^{-3}$}$ and $ L_{*} = 
1.1 \times 10^{10} h^{-2} L_{\odot}$ (see also \cite{Efst88}).
Then the mean luminosity density, corresponding to (\ref{eq:lum}), is:
\begin{equation}\label{eq:mlum}
\langle L \rangle = \int L \Phi(L) dL = \phi_{*} L_{*} \Gamma(\alpha + 2) 
\approx 2 \times 10^{8} h L_{\odot} \mbox{ Mpc$^{-3}$}
\end{equation}
Using the value of $\rho_\mathrm{cr}$ (\ref{eq:crden}) we have:
\begin{equation}\label{eq:m/l}
\frac{M}{L} = \frac{\rho_{\circ}}{\langle L \rangle} = \frac{\Omega_\mathrm{m} 
\rho_\mathrm{cr}}{\langle L \rangle} \simeq 1400 \; \Omega_\mathrm{m} \; h \; 
M_{\odot}/L_{\odot} \;.
\end{equation}
In Table 2  we summarize the 
$\langle M/L \rangle$ values found at different 
scales and the corresponding contribution to $\Omega_\mathrm{m}$.

Since galaxy clusters are the deepest potential wells in the Universe
and they accumulate baryonic and DM from large volumes, it is
expected that their $M/L$ ratio could represent 
the Universal value. This view is supported by the fact that
the increasing trend of $M/L$ with scale (seen in Table 2) reaches a
plateau at the corresponding value of the clusters (cf. \cite{BLD95},
\cite{Bac00}). 
Therefore the universal value, as given by the clusters is:
$$\Omega_\mathrm{m} \simeq 0.23^{+0.12}_{-0.06}$$
The analysis of the CNOC sample of 16 distant clusters
 ($0.17<z<0.55$), provides a consistent value but with significantly
 smaller uncertainty; $\Omega_\mathrm{m} \simeq 0.19 \pm 0.06$ 
(see \cite{Carl97} and references therein).
\begin{table}[t]
\centering
\label{tab:om}
\tabcolsep 12pt
\caption[]{Mass-to-light ratios and contribution to
$\Omega_\mathrm{m}$ for the different scales.}
\begin{tabular}{lccc} \hline
Object & scale ($h^{-1}$ Mpc) &$\langle M/L \rangle \; h^{-1}$ &
$\Omega_\mathrm{m}$  \\ \hline
Spirals & 0.02 & $10\pm 2$ & 0.0071 $\pm 0.0015$ \\
Ellipticals & 0.02 & $25\pm 5$ & 0.018$\pm 0.004$ \\
Galaxy pairs & 0.1  & $80\pm 20$ & $0.057\pm 0.012$ \\
Groups & 0.8  & $180\pm 60$ & $0.13\pm 0.09$ \\
Clusters & 1.5 & $320^{+170}_{-85}$ & $0.23^{+0.12}_{-0.06}$ \\ \hline
\end{tabular}
\end{table}

\subsection{Cluster Baryon Fraction:}
Assuming that on the scales from which clusters accrete
matter during their formation, there is no segregation of baryons and
DM, then the ratio of baryonic to total matter 
in clusters is representative of the universal value 
($\Omega_\mathrm{B}/\Omega_\mathrm{m}$),
something supported also by hydro-dynamical numerical simulations.
Galaxies add up to only $\sim 5\%$ of the total cluster mass, while
the hot gas, which fills the space between galaxies, accounts for
$\sim 20\%$.

So adding up the galactic and gas contribution to the mass of the
cluster we obtain a measure of the total baryonic cluster mass,
$M_\mathrm{B}$ and then if we can measure the total mass of the cluster,
$M_\mathrm{tot}$, we can estimate $\Omega_\mathrm{m}$.

Assuming hydrostatic equilibrium, the gas traces the cluster total
mass, and using (\ref{eq:Tgr}) we can obtain, $M_\mathrm{to}$.
Most studies (cf. \cite{WNEF} \cite{Mohr} \cite{EF99}) find:
\begin{equation}
\frac{\Omega_\mathrm{B}}{\Omega_\mathrm{m}}=
\frac{M_\mathrm{B}}{M_\mathrm{tot}} \simeq 0.10-0.13
\;\;\;\; \mbox{\rm for $h=0.72$}
\end{equation}
From the primordial nucleosynthesis constraints (see \ref{eq:BBN}) and 
from the recent {\sc Bommerang} results we have that 
$\Omega_\mathrm{B} h^{2} \simeq 0.020 \pm 0.004$), which for $h=0.72$ gives
$\Omega_\mathrm{B} \simeq 0.04$ and therefore:
$$
\Omega_\mathrm{m} \simeq 0.35 \; (\pm 0.05)
$$
Note however that in a recent study \cite{SBl01}, the application of various
corrections to account for the clumping of gas and 
the gas fraction gradients, within
the virial radius of a cluster, resulted in
significantly lower values of the
baryon fraction and thus higher values of $\Omega_\mathrm{m}$.

\subsection{Large-Scale Velocity Field:}
As already discussed in section 2, the local gravitational field
produces peculiar velocities superimposed on the general 
expansion (\ref{eq:hub1}). 
Measurements of peculiar
velocities can provide direct information on the mass content of the
Universe, since they can be related to the density fluctuation field,
which itself can be observed directly. The basic idea is 
that a body of mass $M$ will produce a different
gravitational field if it is embedded in a low or high density
Universe. In a high density Universe it will correspond to a lower density
fluctuation than in a low density Universe and thus it will produce
weaker/stronger gravitational effects, respectively.

If $\delta(\vec{x})=(\rho(\vec{x})-\bar{\rho})/\bar{\rho}$ 
is the mass fluctuation at $\vec{x}$ then using
linear perturbation theory, continuity, Euler and Poisson
equations, we obtain the that the growing mode of the
evolution of fluctuations is:
$$\delta \propto D(t) \Longrightarrow
\frac{\dot{\delta}}{\delta}=\frac{\dot{D}}{D}$$
Note that in the EdS universe we have $D(t)=t^{\frac{2}{3}}$. 
Linearizing the mass continuity equation, we obtain:
\begin{equation}\label{eq:con0}
\div\vec{v} = - R \dot{\delta} =R \delta \frac{\dot{D}}{D}
\end{equation}
which has solution:
\begin{equation}\label{eq:lpt}
\vec{v}(\vec{x})=\frac{\Omega_\mathrm{m}^{0.6}}{4 \pi b} \int 
\frac{\vec{x} - \vec{x'}}{|\vec{x} - \vec{x'}|^3} \; 
\delta_\mathrm{tr}(\vec{x}) \; \D^{3} x
 \equiv \frac{\Omega_\mathrm{m}^{0.6}}{b} \vec{D}(\vec{x})
\end{equation}
where the the dipole vector $D(x)$ is related to the gravitational
acceleration vector and $b$ is the bias factor relating the mass-tracer 
fluctuations ($\delta_\mathrm{tr}$) 
with the underline mass fluctuation field (\ref{eq:bias}).
Note that in most cases it is expected that $b>1$, ie., 
the {\it tracer} distribution is more clustered than the 
matter. This has been born out from the study of random Gaussian
fields in which the higher the density fluctuations
the more clustered they are (cf. \cite{Bar86}) and from the
correlation function analysis of extragalactic objects by which it was found
that the {\em relative} bias factor of {\sc abell} clusters, optical
and {\sc iras}
galaxies is [$b_\mathrm{cl}:b_\mathrm{op}:b_\mathrm{IR}=
4.5: 1.3: 1$] (see \cite{PDo94}). It is therefore expected that this
hierarchy of decreasing correlations should continue to the underlying
mass distribution.
 Note that (\ref{eq:lpt}) can be written as \cite{Pee80}:
\begin{equation}
\vec{v} = \frac{2}{3} \frac{\vec{g}}{H_{\circ} \Omega_\mathrm{m}^{0.4}}
\end{equation}
One then needs good estimates of peculiar velocities, knowledge of the matter
tracer fluctuation field to estimate $D(x)$, 
and an understanding of the biasing between matter and light
in order to put constraints on $\Omega_\mathrm{m}$. 

\subsubsection{Local Group Dipole:}
The above test was first applied to the Local
Group, since from the CMB dipole we have an excellent measurement of
its peculiar velocity. 
The dipole moment, $D(x)$, is measured using different
populations of extragalactic objects ({\sc iras}, optical galaxies, AGN's,
{\sc abell/aco} optical or X-ray clusters), weighting each object by $r^{-2}$
or by a weight proportional to $r^{-2}$ (like flux, diameter$^{2}$).
Under the assumption that light traces the mass, then the dipole
moment is a measure of the peculiar force acting on the LG.

In linear theory the peculiar velocity is parallel to the acceleration
and therefore finding an approximate alignment of the two vectors tells us
that the fluctuations causing this motion are present within the depths of the 
sample, provided that 
the dipole moment converges to its final value
before the characteristic depth of the sample. Furthermore, it tells
us that possible local non-linear effects do not strongly affect the
$\vec{D}(x)$ determination.

The multipole components of some mass-tracer distribution are calculated 
by summing moments. For example the monopole and dipole terms are:
$M = 1/4\pi \sum w_\mathrm{i}$ and $\vec{D} = 3/4\pi \sum w_\mathrm{i} 
\vec{r}_\mathrm{i}$ 
where $w_\mathrm{i} \propto r^{-2}$.
The dipole vector, $\vec{D}$, is calculated by weighting the unit directional 
vector pointing to the position of each galaxy, with the weight 
$w_\mathrm{i}$, of that
galaxy and summing over all $N$ available galaxies in the survey.
Note however, that even a uniform distribution would produce a dipole, the
so-called shot-noise dipole, if it is sparsely sampled. 
The shot noise error is
$\langle \vec{D} \cdot \vec{D} \rangle^{\frac{1}{2}} = 
3/4\pi \; N^{\frac{1}{2}} \langle w^{2} \rangle^{\frac{1}{2}}$,
and even in a clustered distribution, the estimated dipole will always
have such a shot-noise contribution which should be taken into
account.
A further difficulty is that whole-sky distributions of
extragalactic mass-tracers are unavailable, either due to survey limitations,
extinction near the Galactic plane, cirrus emission, magnitude, flux
or diameter limits (which will cause a different population mix to be
sampled at different depths, and the contributions of the `faint' 
or `small' objects to be missed). Therefore, complicated corrections
should be applied, which usually have well understood properties and
in any case are always tested with numerical or Monte-Carlo
simulations.
\begin{figure}[t]
\centering{
\mbox{\epsfxsize=16cm \epsffile{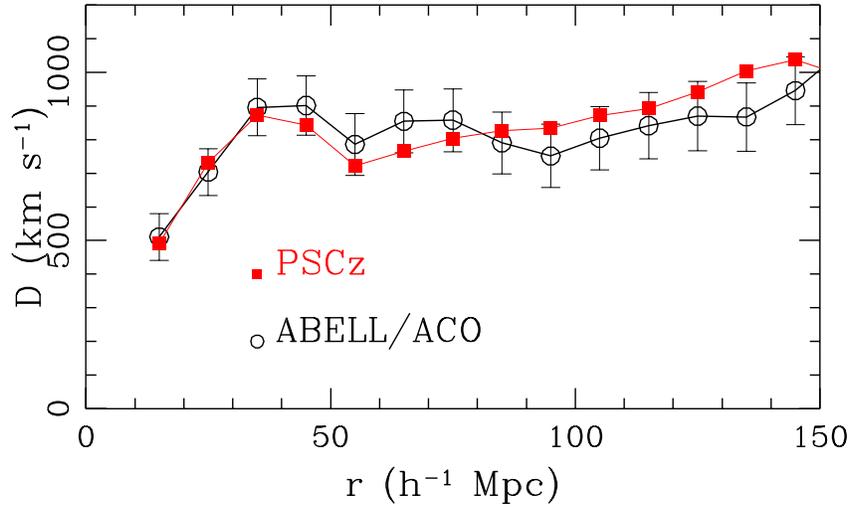}}}
\caption{Dipole amplitude build up as a function of distance of the
{\sc iras pscz} galaxy and the {\sc abell/aco} cluster samples. The
{\sc abell/aco} cluster dipole has been scaled down by a factor of $\sim 4$,
to take into account the relative bias factor (from \cite{Pli00}).}
\end{figure}

Results from many different galaxy catalogues (optical or IR)
tracing depths up to $\sim 100 - 120 \;h^{-1}$ Mpc, show dipoles that 
are well aligned with the CMB dipole
(mass dipole) which indicates that galaxies 
trace the mass distribution. Some recent analysis of {\sc iras} galaxies
provide values:
$$\Omega_\mathrm{m} \approx 0.62 (\pm 0.13) \; b_\mathrm{IR}^{1.66} $$
(cf. \cite{SW95}, \cite{RR00} and references therein) and
$$\Omega_\mathrm{m} \approx 0.55 (\pm 0.2) \; b_\mathrm{op}^{1.66} $$
(cf. \cite{Hud93}).
The difference between the above values is most probably 
due to the different biasing factors ie.,
$b_\mathrm{IR} < b_\mathrm{op}$, which 
is natural since optical galaxies trace more accurately the 
deep potential wells (clusters) while
the {\sc iras} galaxies trace better the field. In fact it has been found that
$b_\mathrm{op}/b_\mathrm{IR} \simeq 1.2-1.4$ (cf. \cite{PDo94}
\cite{Bak98} \cite{Pli00} and references therein).
Furthermore, if IR galaxies are biased with respect to mass (ie., if
$b_\mathrm{IR} \gtrsim  1.2$) then
the above results could imply: $\Omega_\mathrm{m} \simeq 1$.
 
Now, {\em galaxy clusters}, being the largest gravitationally-collapsed 
structures in the universe and luminous enough to be detected to very
large distances, have also been used to probe the local acceleration field.
Most studies are based on the optically selected {\sc abell/aco}
clusters \cite{ab1} and they provide strong
evidence that the LG dipole has significant contributions from depths
up to $\sim 160h^{-1}\,$Mpc (see Fig.12 and \cite{SVZ91}, 
\cite{PV91}, \cite{BP96}).
However, due to the the volume incompleteness of richness class R=0 clusters 
(cf. \cite{PW92}) and to optical projection effects (enhancement of 
galaxy density along the direction of foreground rich clusters which cause 
inherently poor background clusters or groups to appear rich enough to be
included in the sample), these results had to be checked. Using
well defined X-ray cluster samples free of the above effects \cite{Ebel}
indeed the results were verified \cite{PK98}. 
The results based on optical or X-ray clusters, imply:
$$\Omega_\mathrm{m} \approx 0.07 - 0.09 \; b_\mathrm{cl}^{1.66} $$
Note that from the correlation function analysis of {\sc abell/aco} clusters
\cite{PDo94} 
we have that the relative bias factor between clusters and {\sc iras}
galaxies is around $b_\mathrm{cl,IR} \simeq 4.5$ and if the bias factor
of {\sc iras} galaxies with respect to the 
mass is $b_\mathrm{g}\ge 1$, then the
above results are consistent with $\Omega_\mathrm{m} \simeq 1$.

It is extremely interesting that the galaxy distribution in the local universe 
(within $\sim 100 \;h^{-1}$ Mpc) produces a gravitational acceleration that is 
aligned with that produced by the matter distribution (as determined by the 
CMB dipole) and in the same time the distribution of clusters of galaxies on 
larger scales $R \lesssim 250 \; h^{-1}$ Mpc produce also a
gravitational acceleration aligned with that of the matter
distribution. Furthermore the galaxy and cluster 
distributions produce Local Group acceleration profiles 
that are directly proportional to each other (see Fig.13),
while the dipole in equal volume shells, seems to
be roughly aligned with the CMB dipole direction out to very large
depths (cf. \cite{PV91}, \cite{BP98}).
This implies that there is a coherent anisotropy in the mass 
distribution 
over a region with a diameter of $\sim 300$ $h^{-1}$ Mpc, which then
sets a
lower limit to the scales over which the Cosmological Principle applies.

However, there is a dichotomy among different studies trying to
identify the convergence scale of the dipole, or equivalently what is
the largest scale over which we observe bulk motions. The previously
discussed dipole studies as well as some peculiar velocity studies
(cf. \cite{LP94}, \cite{Hud99}, \cite{Wil99}),
support the view of a large convergence depth (radius of $\sim
150$ $ h^{-1}$ Mpc). Other peculiar velocity studies (cf. 
\cite{Cour00}, \cite{DG00}, \cite{D99}, \cite{Co00}, \cite{DaCo})
support a significantly smaller convergence depth $\sim 60$
$h^{-1}$ Mpc.

\subsubsection{POTENT - from Radial Velocities to Density Field:}
In the previous analysis only one velocity was used, that of the
Local Group, and although it is very well measured we still have the
problem of cosmic variance. Therefore, ideally the
velocity-acceleration comparison should be performed for a number of
``observers''. This has been possible due to the {\sc POTENT} algorithm
proposed in \cite{Bet89} and developed extensively by
Dekel and his collaborators (cf. \cite{Dek99} and references therein).

The basic idea follows.
The large-scale velocity field, evolving via gravitational
instability, is expected to be irrotational $\rotv=0$.
This remains a good approximation in the
mildly-nonlinear regime as long as the field is properly smoothed. 
This implies that the velocity field can be derived from a
scalar potential,
$$\vv(\vec{x})=-\vnabla\Phi(\vec{x})\;,$$
and thus the potential can in principle 
be computed by integration along the lines of sight,
\begin{equation} 
\Phi(\vec{x}) = -\int_0^r u (r',\theta,\phi) \D r' \;.
\end{equation}
The two missing transverse velocity components are then recovered by
differentiation. Then from (\ref{eq:con0})
we recover the density fluctuation field, which can then be compared
to the observed density field, determined from large whole-sky surveys.
The current sampling of galaxies enables reliable dynamical analysis,
with a smoothing radius as small as $\sim\!10\hmpc$, where
$\vert\div\vec{v}\vert$ 
obtains values larger than unity and therefore
mildly non-linear effects play some role.
\begin{figure}[t]
\centering{
\caption{(Included separately as a JPG image)
Comparison of predicted density and velocity fields: Left
panel shows the observed {\sc abell/aco} density field and the corresponding
predicted velocity field. Right panel shows the matter density field
predicted by {\sc POTENT} and the peculiar velocity field of 
Mark III galaxies (from \cite{BZPD}).}
}
\end{figure}

The most reliable density-density analysis, incorporating 
certain mildly non-linear corrections, is the
comparison of the {\sc iras} 1.2 Jy redshift survey and the Mark~III
catalogue of peculiar velocities yielding, at Gaussian smoothing of 
$12\hmpc$ \cite{sig98}:
$$\Omega_\mathrm{m} \simeq 0.82 (\pm 0.16) \; b_\mathrm{IR}^{1.66} $$
A similar analysis, using optical galaxies \cite{Hud95}
has provided a somewhat lower value:
$$ \Omega_\mathrm{m} \simeq 0.6 \; (\pm 0.15) \; b_\mathrm{op}^{1.66} $$
in accordance with the expected higher biasing parameter of optical
galaxies with respect to {\sc iras} ones. These results are consistent with
the dipole analyses and with
$\Omega_\mathrm{m} \simeq 1$ for $b_\mathrm{IR} \gtrsim 1.2$.

However, a variety of methods using {\em v-v} comparisons 
(eg. {\sc VELMOD} - \cite{WS98}),
developed to compare observed and derived
velocities (using either the {\sc iras} or {\sc ors} gravity fields),
typically yield values of:
(cf. \cite{Wil00} and references therein, \cite{Br01}, \cite{Blak}):
$$\Omega_\mathrm{m} \simeq 0.3 \; (\pm 0.1) \;
b_\mathrm{IR}^{1.66}$$
$$\Omega_\mathrm{m} \simeq 0.14 \; (\pm 0.05) \;
b_\mathrm{op}^{1.66}$$
which are consistent with $\Omega_\mathrm{m} <1$ for any reasonable value
of $b_\mathrm{g}$.
Therefore, there seems to be a discrepancy between different analyses,
even if in some cases, they use the same data, a fact that needs
further study and tests of the reliability of each method.

A study \cite{BZPD} using the {\sc abell/aco} clusters to trace
the density field and comparing it with the {\sc potent} reconstructed
field from the Mark III catalogue of peculiar velocities (see Fig.14)
found:
$$\Omega_\mathrm{m} \approx 0.07 - 0.09 \; b_\mathrm{cl}^{1.66}$$
in good agreement with the dipole analysis of {\sc abell/aco} clusters and
consistent with $\Omega_\mathrm{m} \simeq 1$ for the estimated value
of $b_\mathrm{cl}$ \cite{PDo94}.

\subsubsection{Local Group infall to Virgo:}
This is an interesting method to calculate $\Omega_\mathrm{m}$ on scales of 
$\sim 10$ $h^{-1}$ Mpc.
One relates the Local Group infall (towards the centre of the Local 
supercluster) velocity with the acceleration induced to the LG by the mass 
overdensity in the Local Supercluster, assuming a point-mass approximation.
We have from (\ref{eq:lpt}):
\begin{equation}\label{eq:mmm}
\vec{v}_\mathrm{in} = \frac{2}{3} 
\frac{\vec{g}}{H_{\circ} \Omega_\mathrm{m}^{0.4}} =
\frac{2}{3} \frac{G \delta M}{H_{\circ} \Omega_\mathrm{m}^{0.4} r^{2}}
\end{equation}
From (\ref{eq:bias}) we have that the galaxy fluctuations 
is a biased tracer of the underline mass fluctuation field:
\begin{equation}
b \frac{\delta M}{M} \simeq \delta_\mathrm{g} \Longrightarrow
\delta M \simeq \frac{M \delta_\mathrm{g}}{b} \;.
\end{equation}
Thus from (\ref{eq:mmm}) we have:, 
$$v_\mathrm{in} = \frac{1}{3} \frac{\Omega^{0.6}_\mathrm{m}}{b} \; cz \;
\delta_\mathrm{g} (1+\delta_\mathrm{g})^{-1/4}$$
where the mildly non-linear correction on the right-hand side is 
according to \cite{Yah}. A recent study \cite{Tonry} 
using the SBF method to determine the local velocity
field within $cz \lesssim 3000$ km/sec find a Virgo-centric infall of
$v_\mathrm{in} \sim 140$ km/sec in agreement with $\sim 160$ km/sec,
implied from the Virgo contribution to the X-ray cluster dipole
\cite{PK98}. Furthermore we have that
$cz \simeq 1005$ km/sec \cite{MAH} \cite{ST95}
and $\delta_\mathrm{g} \simeq 2.8 \pm 0.5$. Therefore we obtain:
$$\Omega_\mathrm{m} \simeq 0.1 (\pm 0.02)\; 
\left(\frac{v_\mathrm{in}}{150 \;\mbox{km/sec}}\right)^{1.66} \;
b_\mathrm{op}^{1.66} $$
Although this method of determining $\Omega_\mathrm{m}$ is `clean',
the fact that the local peculiar velocity field is affected by
mass concentrations well beyond the Local supercluster introduces a further 
uncertainty in the determination of $\Omega_\mathrm{m}$. 

\subsubsection{Velocity-Field results:}
The outcome of the different large-scale dynamical studies do not
converge to a unique value of the mass density parameter. There is
need to check the methods and understand the source of this
discrepancy. In Table 3 I sum up the different results from the
different velocity-field analyses.
\begin{table}[t]
\centering
\tabcolsep 12pt
\caption[]{Results from some recent velocity field studies}
\begin{tabular}{lccc} \hline
Type of study & $\Omega_\mathrm{m} \; b^{-1.66}_\mathrm{IR}$ 
&$\Omega_\mathrm{m} \; b^{-1.66}_\mathrm{op}$ & 
$\Omega_\mathrm{m} \; b^{-1.66}_\mathrm{cl}$ \\ \hline
Dipole & 0.62 & 0.55 & 0.08 \\
POTENT/Mark III & 0.82 & 0.6 & 0.08 \\
VELMOD & 0.3 & 0.14 & \\
LG-infall &  & 0.1 & \\ \hline
\end{tabular}
\end{table}

\subsection{Rate of Cluster Formation Evolution:}
The rate of growth of perturbations is different
in universes with different matter content.
For example, 
the perturbation growth in a $\Omega_\mathrm{m}=1$ universe is
proportional to the scale factor, ie., $\delta \propto (1+z)^{-1}$,
while in the extreme case of an empty universe ($\Omega_m =0$); $\delta
=$ constant. From (\ref{eq:omeg_z}) we see that 
$\Omega_\mathrm{m}<1$ universes 
will behave dynamically as an $\Omega=1$ universe at large
enough redshift, and at some redshift $z\sim 1$ 
curvature dominates and
perturbations stop evolving and freeze, allowing 
clusters to relax up to the present epoch much more than in an
$\Omega_\mathrm{m}=1$ model, in which clusters are still forming.
This can be seen clearly in Fig.15 where 
we plot the evolution of the perturbation growth factor, defined as:
$$f = \frac{\D \ln \delta}{\D \ln R} \;.$$
For a $\Omega_\mathrm{\Lambda}=0$ universe, $f\simeq \Omega_\mathrm{m}^{0.6}$
(cf. \cite{Pee93}) and thus for the EdS model $f=1$. For 
$\Omega_\mathrm{\Lambda}> 0$ models 
there is a redshift dependence of $f$, but in the present epoch it is
indistinguishable from the corresponding value of the open
($\Omega_\mathrm{m}=1-\Omega_\mathrm{\Lambda}$) model \cite{Lah91}.
It is evident that an $0<\Omega_\mathrm{\Lambda}<1$
universe behaves as an $\Omega_\mathrm{m}=1$ model up to a 
lower redshift than
the corresponding open model, while at redshifts $z\lesssim 1$ it
behaves like an open model, which implies that clusters
should be dynamically older in such a model than in the EdS.
\begin{figure}[t]
\centering
\mbox{\epsfxsize=12cm \epsffile{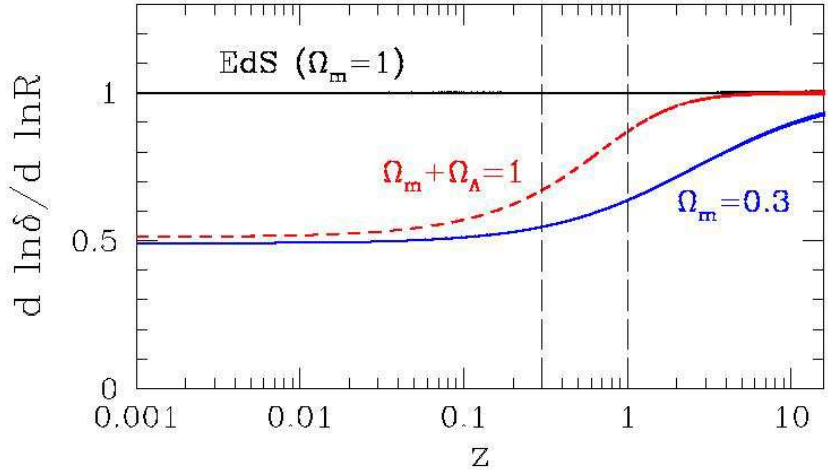}}
\caption{The evolution of the perturbation growth factor, $f$, in 3
models (EdS, open $\Omega_\mathrm{m}=0.3$ and the currently popular flat
$\Omega_\mathrm{\Lambda}=0.7$). The vertical dashed lines indicate the redshift
range $0.3\lesssim z \lesssim 1$ where the 3 models, jointly,
differ maximally.}
\end{figure}
Therefore one should be able to
put constraints on $\Omega$ from the evolution of various
indicators of cluster formation, especially in the range where
the dynamical evolution between the models differs maximally (vertical
dashed lines in Fig.15). 

Ideally, one would like to
study the evolution of the cluster mass function but since light is
what we observe (temperature as well - due to the hot ICM X-ray
emission), various related indicators are usually studied
(Luminosity function, temperature function, morphology etc), but then
one has to pass through the machinery provided by the
Press-Schechter formalism \cite{PS74},
which gives the mass function of collapsed
halos at any epoch as a function of the cosmological parameters that
enter through the assumed power spectrum of perturbations.

\subsubsection{Luminosity function:} 
Based mostly on {\sc Einstein} and {\sc Rosat} surveys, many
studies have found an evolving X-ray luminosity function,
ie., less $z\gtrsim 0.3$ clusters than
expected for a no-evolving luminosity function, ie., a negative evolution 
(cf. \cite{gioia01} and references therein). Such a
behaviour is expected in models with 
$$ \Omega_\mathrm{m} \simeq 1$$
However, see \cite{Hen02} for a different view.
\subsubsection{Temperature function:} Estimates of the temperature of the
X-ray emitting ICM gas can be reliably estimated from the iron
line-emission. Then the cluster temperature can be either transformed to a
mass (assuming hydrostatic equilibrium and isothermality) and thus
derive a mass function to compare with the Press-Schechter predictions
(cf. \cite{RB02} and references therein) pointing to
$\Omega_\mathrm{m}<0.3$ ,
or use the evolution of the temperature distribution function. 
Again different studies find either no evolution
(cf. \cite{eke}, \cite{Hen00}) pointing to 
$$\Omega_\mathrm{m} \simeq 0.3 - 0.5$$
or evidence for evolution \cite{VL99} \cite{Blan00}
pointing to
$$\Omega_\mathrm{m} \simeq 0.7 - 1$$
\subsubsection{Evolution of $L-T$ relation:}
Under the assumption of hydrostatic equilibrium and isothermality one can
easily show, from Euler's equation, that the bremsstrahlung radiation
temperature is $T\propto M_\mathrm{v}/R_\mathrm{v}$ 
(where $M_\mathrm{v}$ and $R_\mathrm{v}$ are the cluster
virial mass and radius). Using the spherical collapse top-hat model 
\cite{Pee80} one
obtains $R_\mathrm{v} \propto  T^{1/2} \; \Delta(z)^{-1/2} \; E(z)^{-1/2}$, and
then by using (\ref{eq:X1}):
\begin{equation}\label{eq:X-more}
L_\mathrm{x} \propto f^{2}_\mathrm{g} M_\mathrm{v}^{2} T^{1/2} 
R_\mathrm{v}^{-3}
\end{equation}
where $f_\mathrm{g}$ is the gas mass fraction.
Then one finds (cf. \cite{BN98}):
$$L_\mathrm{x} \propto f^{2}_\mathrm{g} \; T^{2} \; 
\Delta(z)^{1/2} \; E(z)^{1/2}$$
where $E(z)$ is given by (\ref{eq:new_fre}) and $\Delta(z)$ is the
ratio of the average density within the virialized cluster 
($\lesssim R_\mathrm{v}$) and the critical density at redshift $z$, which also
depends on the cosmological model. However, this model fails to
account for observations which show
a steeper $T$-dependence, $L_\mathrm{x} \propto T^{3- 3.3}$ 
(such a dependence
can be recovered from (\ref{eq:X-more}) if $f_\mathrm{g} \propto T^{1/2}$).
In any case, the $L-T$
relation is expected to evolve with time in a model dependent
way. Most studies (see references in \cite{Sch01}) have
found {\em no} evolution of the relation while a recent
study of a deep ($z \sim 0.85$) 
{\sc Rosat} cluster survey \cite{Borg01} found: 
$$\Omega_\mathrm{m} \sim 0.35$$ 
However, there are many physical mechanisms that affect this relation 
(eg. gas cooling, supernova feedback etc)
and in ways which are not fully understood (cf. \cite{Gov02}).
\subsubsection{Evolution of Cluster Morphology:} 
As we have already discussed, in an open or a flat with
vacuum-energy contribution universe it is expected that
clusters should appear more 
relaxed with weak or no indications of substructure. 
Instead, in a critical density model, such systems continue to form even today
and should appear to be dynamically active 
(cf. \cite{Ri}, \cite{Ev}, \cite{Lac}).
Using the above theoretical expectations as a cosmological tool is
hampered by two facts (a) {\it Ambiguity in identifying cluster substructure}
(due to projection effects) and (b) {\it Post-merging relaxation time 
uncertainty} (cf. \cite{Sar01}). However, criteria of recent merging
could be used to identify the rate of cluster morphology evolution and
thus put constraints on $\Omega_\mathrm{m}$. 
Such criteria have been born out of numerical simulations
(cf. \cite{Roe93}, \cite{Roe99}) 
and are based on the use of 
multiwavelength data, especially optical and X-ray data but radio as
well (cf. \cite{Zab}, \cite{Sch99}).
The criteria are based on the fact that gas is collisional while galaxies are
not and
therefore during the merger of two clumps, containing galaxies and gas,
we expect: ({\it 1}) a difference in the spatial positions of the 
highest peak in the galaxy and gas distribution,
({\it 2}) due to compression, the X-ray emitting gas to be elongated 
perpendicularly to the merging direction, and
({\it 3}) temperature gradients to develop due to
the compression and subsequent shock heating of the gas. The first two
indicators are expected to decay within $\sim1$ Gyr after the
merger, while the last may survive for a considerably longer period
(see for example Fig.16).
\begin{figure}[t]
\mbox{\epsfxsize=6cm \epsffile{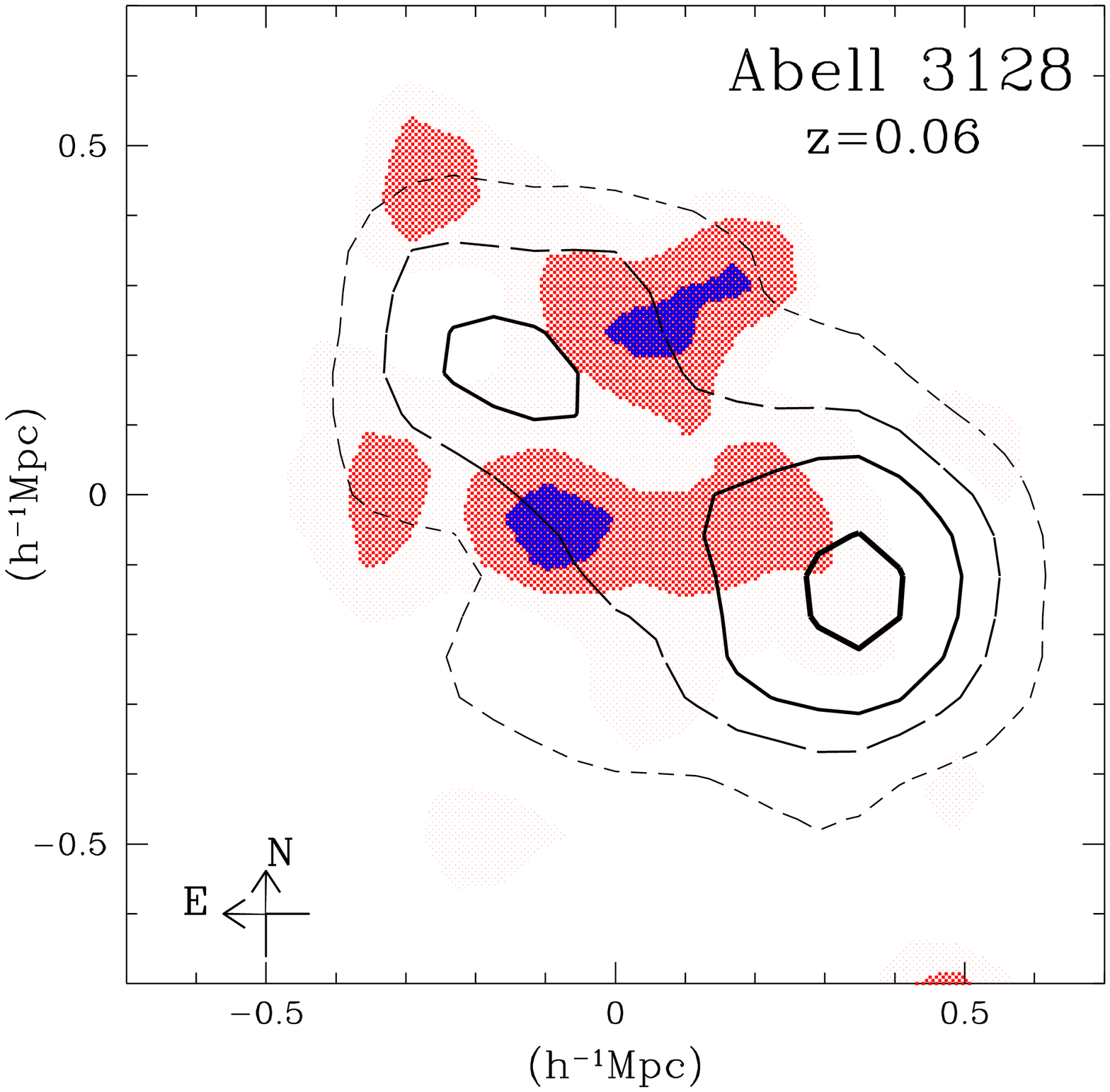}}
\hfil
\mbox{\epsfxsize=6cm \epsffile{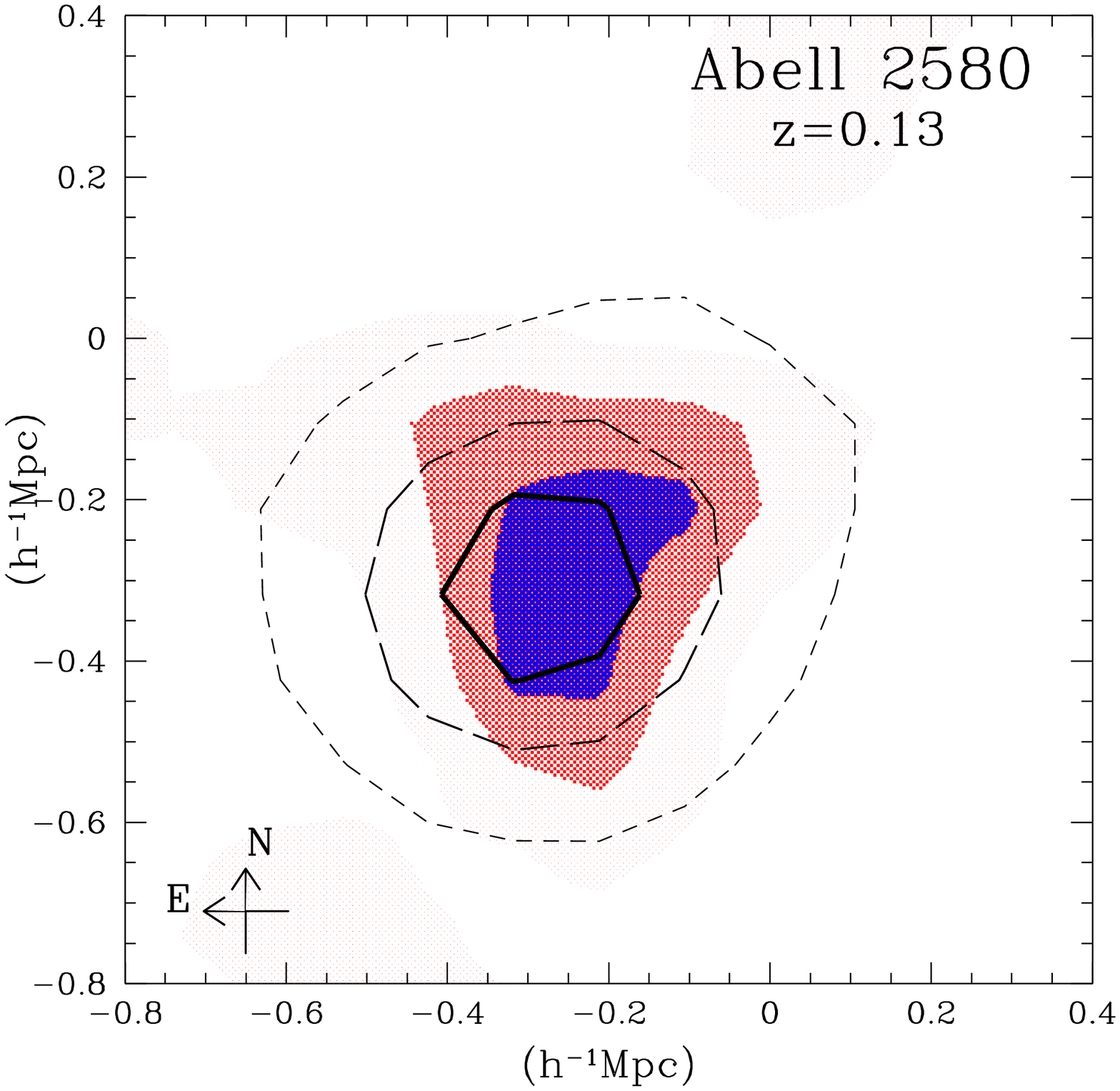}}
\caption{Optical {\sc apm} (colour) and {\sc Rosat} X-ray (contour) images
of 2 {\sc abell} clusters. Peaks of the {\sc apm} galaxy distribution is shown 
in blue. A3128 has the signature of a recent merger: the peaks
in the distribution of galaxies and in X-ray emitting gas are
orthogonal to each other. A2580 on the other side seems a smooth
relaxed cluster with the gas and galaxies tracing the cluster potential
(from \cite{Pl01}).}
\end{figure} 
For such a study to be fruitful, a large number of clusters, ideally
covering the 
redshift range $0.3 \lesssim x \lesssim 1$, must be
imaged in both the optical and X-ray band.

However, a rather cruder but still useful test of cluster morphological
evolution could be used. For example, cluster ellipticity is a
relatively well defined quantity; although systematic effects due to
projections in the optical or the strong central concentration of the
X-ray emitting gas (since $L_\mathrm{x} \propto n^{2}_\mathrm{e}$), 
should be taken into account (cf. \cite{KBPG}). 
An early study, using the Lick map \cite{PBF91}, had found
that cluster ellipticity decreases with redshift, however due to
possible systematic effects involved in the construction of the data,
they did not attach any weight to this discovery. Recently, two studies using
optical and/or X-ray data \cite{MelC} \cite{Pl02} (see also \cite{Pl01})
found that indeed the cluster ellipticity decreases with redshift 
in the recent past, $z\lesssim 0.15$ (see Fig.17) 
This was interpreted by \cite{MelC} as an
indication of a low-$\Omega_\mathrm{m}$ universe because in such a universe one
expects that merging and anisotropic accretion of matter along
filaments will have stopped long ago. Thus the clusters should be
relatively isolated and gravitational relaxation will tend to isotropize the
clusters reducing their ellipticity, more so in the recent
times. 

If this is the case then one should expect an evolution of the
temperature of the X-ray emitting gas as well as the X-ray cluster
luminosity which should follow the same
trend as the cluster ellipticity, decreasing at recent times, since
the violent merging events, at relatively higher redshifts, will 
compress and shock heat the diffuse ICM gas \cite{Ritchi}.
Such evidence was presented in \cite{Pl02} using a compilation of
measured ICM temperatures and luminosities in two volume limited X-ray
cluster samples (based on the XBAC and BCS samples).
Also, one could 
naively expect an evolution of the cluster velocity dispersion, increasing
at lower redshifts, 
since virialization will tend to increase the cluster
`thermal' velocity dispersion. In \cite{GM01}
no evolution was found between a local sample ($z\lesssim
0.15$) and a distant one $0.15 \lesssim z
\lesssim 0.9$. However, unrelaxed clusters can also
show up as having a high velocity dispersion due to either possible large
peculiar velocities of the different sub-clumps \cite{Rose}
or due to the possible sub-clump virialized nature. 
Therefore, a better physical understanding of the
merging history of clusters is necessary in order to be able
to utilize the velocity dispersion measure as an evolution criterion.
\begin{figure}[t]
\mbox{\epsfxsize=12cm \epsffile{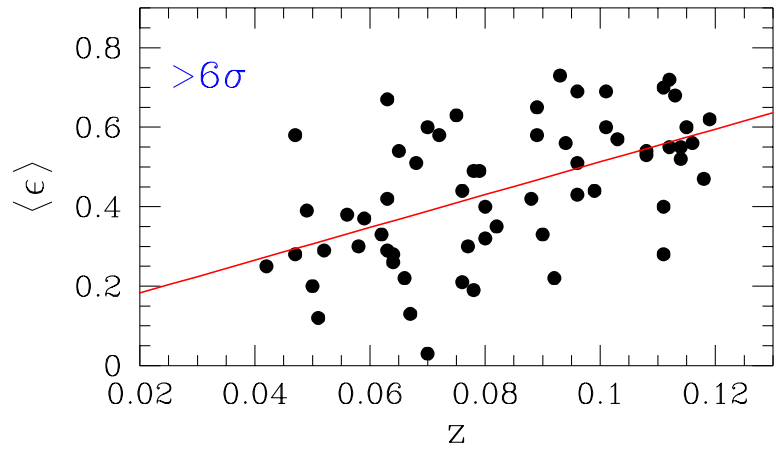}}
\caption{The evolution of ellipticity in {\sc apm} clusters with 
significant substructure \cite{Pl01}}
\end{figure}

Other related studies, using the morphological characteristics of the
large-scale structures, have been used to place cosmological
constraints. For example, the shapes of superclusters and voids, using
the {\sc iras-pscz} 
redshift survey and the {\sc abell/aco} cluster distribution
show a clear preference for a
$\Lambda$-CDM model over a $\Omega_\mathrm{m}=1$ model 
\cite{BPR01} \cite{PB02b} \cite{KBP02} \cite{Mirand} but see \cite{Arb}.

\newpage

\section{Summary}
I have attempted to present the basic ideas behind many (but not all) 
of the current
methods used to estimate various Cosmological parameters, especially the
Hubble constant ($H_{\circ}$), 
the curvature of the Universe ($\Omega$) and its matter content
($\Omega_\mathrm{m}$). To do so I reviewed some basics of the standard,
Robertson-Walker, Cosmology 
in order to highlight the interrelations of the Cosmological
parameters and the way they affect the global dynamics of the Universe.

The results of a variety of different analyses, based on a multitude
of data, point towards a {\em concordance} model, which is a flat, 
$\Omega_\mathrm{\Lambda} \simeq 0.7$, with an inflationary spectral index
$n\simeq 1$, $h\simeq 0.72$, $\Omega_\mathrm{B}\simeq 0.04$, 
$t_{\circ}\simeq 13.1$ Gyr's.
Nevertheless there are conceptual problems and open issues that may
or may not prove to be daunting.

For example, the $\Omega_\mathrm{\Lambda}=0.7$, $h=0.7$ model 
seems to have problems
generating the correct power on galactic scales. Detailed numerical
simulations of \cite{Klypin} show that this model has much more power
on small scales to be reconciled with observations (see discussion in
\cite{Primack}).
Furthermore, there is the fine tuning problem. Why does the energy
density of the vacuum have a value like:
$\Omega_\mathrm{\Lambda} \sim \Omega_\mathrm{m} \sim 1$, implying that
it dominates the Universe just {\sc now}! An easy way out would be to 
invoke {\em anthropic} arguments \cite{Barr}, but it is too feeble
a justification, especially since the different contributions to
$\Lambda$ at the early phase transitions are 50 - 100 orders of
magnitude larger than what is observed. One would have to
invoke a fine-tuning in order for the different contributions to
cancel out, but yet not completely! However, theoretical models are being
developed in an attempt to alleviate such problems
(cf. \cite{Elizalde}), some 
by allowing $\Lambda$ to be a function of time 
(see \cite{Varun} and references therein).

\section*{Acknowledgments}
I would like to thank Spiros Cotsakis and Lefteris Papandonopulos 
for organizing this wonderful school and for their hospitality in the
island of Pythagoras. I would also like to thank the students of the
school for their interest, their stimulating discussions and
their ...dancing abilities. Many thanks to Ed Chapin that had the
patience to go through the text and correct my ``Greekisms''.

\end{document}